\newif\ifsubmode
\submodefalse

\newif\ifprintfig
\printfigtrue

\newif\ifemulate
\emulatetrue
\ifsubmode
\documentclass[12pt,preprint]{aastex}
\received{}
\accepted{}
\journalid{}
\articleid{}
\else
   \documentclass{emulateapj}
   \submitted{{\it Invited Review Article, Accepted for Publication in Annalen der Physik}}
\fi
\usepackage{multirow,color,natbib,wrapfig,ulem}
\begin{document}
\newcommand{\jcap}{Journal of Cosmology and Astroparticle Physics}
\newcommand{\nar}{New Astronomy Reviews}
\title{Re-Examining Astrophysical Constraints on the Dark Matter Model}
\author{Alyson\ M.\ Brooks\altaffilmark{1}} 
\altaffiltext{1}{Department of Physics \& Astronomy, Rutgers, The State University of New Jersey, 136 Frelinghuysen Rd,
Piscataway, NJ 08854; abrooks@physics.rutgers.edu}
\shortauthors{Alyson Brooks}
\begin{abstract}
Recent high-resolution simulations that include Cold Dark Matter (CDM) 
and baryons have shown that baryonic physics can dramatically alter 
the dark matter structure of galaxies.  These results modify our 
predictions for observed galaxy evolution and structure.  Given these 
updated expectations, it is timely to re-examine observational 
constraints on the dark matter model.  A few observations exist that 
may indirectly trace dark matter, and may help confirm or deny possible 
dark matter models.  Warm Dark Matter (WDM) and Self-Interacting Dark 
Matter (SIDM) are currently the favorite alternative models to CDM.  
Constraints on the WDM particle mass require it to be so heavy that 
WDM is nearly indistinguishable from CDM.  The best observational test 
of SIDM is likely to be in the dark matter distribution of faint dwarf 
galaxies, but there is a lack of theoretical predictions for galaxy 
structure in SIDM that account for the role of baryons.
\end{abstract}
\maketitle

\section{Introduction}

There is six times more mass in dark matter than baryonic matter
\footnote[2]{Throughout this review we adopt the common convention of 
astronomers to refer to all standard model particles, including leptons, 
as baryons.} in our Universe \citep{WMAP9,Planck2013}.  For decades it has 
been assumed that, because dark matter is so much more common than baryons, 
dark matter dominates the gravity in the Universe, and that wherever the 
dark matter is, baryons must follow.  This assumption led galaxy theorists 
to make predictions for the formation of galaxies using dark matter only, 
neglecting baryonic physics, despite the fact that galaxies like our own 
Milky Way are baryon-dominated within their inner $\sim$10kpc.  In doing 
so, a number of discrepancies between galaxy formation theory and 
observations were identified, particularly on ``small scales,'' i.e., in 
small galaxies and in the central regions of galaxies \citep{PrimackAnnalen}.  
To address these problems, alternative models to the standard Cold Dark 
Matter (CDM) have been explored.  Recent interest has surged in Warm Dark 
Matter (WDM) and Self-Interacting Dark Matter (SIDM) models as the favorite 
alternatives to CDM amongst galaxy theorists.

However, there has also been a recent reconsideration of the importance 
of baryonic physics in solving CDM's small scale problems.  Observationally, 
it is clear that energy feedback from stars and black holes operates to 
alter the evolution of galaxies.  For example, the existence of gas outflows 
(``winds'') from galaxies seems to be ubiquitous at high redshift 
\citep{Veilleux2005}.  
Energetic feedback from stars (in the form of radiation pressure from young, 
massive stars, momentum injection by the winds of the same stars, and supernovae) 
has long been included in galaxy simulations, but only recently have simulations 
achieved sufficiently high resolution to deposit this feedback in localized 
regions.  Localized feedback dramatically impacts the evolution of the galaxy 
\citep{Governato2010, Guedes2011, Christensen2012, Agertz2013, Aumer2013, 
Hopkins2013}, and drives the ubiquitous winds that we observe. 

The processes that drive galaxy winds also have a dramatic impact on the 
dark matter structure of galaxies.  Feedback from stars can push the dark 
matter out of the central $\sim$kpc by generating a repeated fluctuation 
in the potential wells of galaxies \citep{Navarro1996, Read2005, deSouza2011, 
Pontzen2012, Teyssier2013, diCintio2014}.  
This result reconciles the dark matter density profile predicted in CDM 
that is steeply rising toward the center \citep[``cuspy,''][]{Navarro1997a, 
Springel2008, Navarro2010} with observations  which instead prefer a 
shallower density slope or even a constant dark matter density ``core''
\citep{vandenbosch2000, deblok2001, deblok2002, Simon2003, Swaters2003, 
Weldrake2003, Kuzio2006, Gentile2007, Spano2008, Trachternach2008, 
deblok2008, Oh2011}. Hence, one of the seemingly intractable problems 
plaguing CDM theory is now thought to be potentially solved by a careful 
consideration of the impact of baryonic physics \citep{Pontzen2014}.

Galaxy winds also solve another problem within CDM galaxy formation theory: the 
existence of bulgeless disk galaxies.  Galaxies are thought to obtain their 
angular momentum through large-scale tidal torques \citep{Peebles1969, 
White1984, Barnes1987, Quinn1992}.  Gas and dark matter start with the 
same angular momentum distribution \citep{vdBosch2002}, with a tail 
of low angular momentum material that is expected to settle at the center 
of galaxies.  Low angular momentum gas should go on to form stars, forming 
large bulges, at odds with observed small or non-existent stellar bulges 
\citep{vdBosch2001, Dutton2009b}. Galaxy winds naturally arise from the 
region where most star formation is occurring, in dense galaxy centers 
where low angular momentum gas resides. Hence, winds naturally drive low 
angular momentum material from galaxies \citep{Brook2011}, and can create 
bulgeless disk galaxies \citep{Governato2010, Teyssier2013}, solving another 
of CDM's small scale problems.

One of the oldest problems and one of the newest problems facing CDM galaxy 
formation theory both relate to the satellites that orbit around our Milky Way 
galaxy.  First, simulations predict that there should be many more satellites 
than we observe \citep{Moore1999, Klypin1999}.  Many of these satellites are expected 
to be ``dark,'' unable to have formed stars due to photoevaporation of their gas 
when the Universe was re-ionized \citep{Quinn1996, Thoul1996, Barkana1999, Gnedin2000, 
Okamoto2008}, though this process alone may not be enough to bring the predicted 
number of massive, luminous satellites into agreement with observations 
\citep{Brooks2013}.  Furthermore, even if we could get the number of satellites 
correct, there exists a population of satellites in simulations run without 
baryons that are much more dense than we observe \citep{Boylan-kolchin2011, 
Boylan-kolchin2012, Tollerud2012, Collins2014}.  This latter problem is also 
known as the ``Too Big to Fail'' problem because the simulated satellites are 
too massive to have failed to form stars, yet we do not observe them.  Again, 
recent high resolution simulations have shown that baryonic effects may reconcile 
both of these predictions with observations \citep{Zolotov2012, diCintio2013, 
Arraki2014}.  The primary physics at work is the fact that gas, unlike dark matter, 
can cool.  In a simulation with baryons, this cooled component adds more mass to the 
center of the parent halo, creating stronger tidal forces that strip mass from the 
satellite galaxies \citep{Penarrubia2010}. This enhanced tidal stripping reduces the 
mass of the satellites, bringing the kinematic predictions in line with the 
observational data \citep{Brooks2014}. The presence of the disk in a baryonic 
simulation (which doesn't exist in a dark matter-only run because dark matter 
cannot dissipate) will also fully destroy roughly 1/3 of the most massive 
satellites, reducing the number of luminous, surviving satellites so that 
it is consistent with observations \citep{Brooks2013, Brooks2014}.  

The lesson we have learned from these studies is that baryons have 
the potential to alter our expectations for the structure of dark matter 
halos that form within CDM.  While CDM does an excellent job of describing 
the large scale structure of the Universe \citep{Hlozek2012}, we can no 
longer neglect the influence of baryons when considering small scales.
It is important to note that, of the problems listed above, the creation 
of bulgeless galaxies cannot be solved by any correction to the dark matter 
model.  Only baryonic feedback is able to explain the loss of low angular 
momentum baryons from galaxies.  

The fact that galaxy winds offer a single, unified solution to the existence 
of both bulgeless disks and dark matter cores is tantalizing evidence that 
these two problems are intimately tied together. Despite this, modifications 
to the dark matter model are still being pursued as another possible explanation 
for the existence of dark matter cores.  Given the fact that baryonic physics 
cannot be neglected (and is in fact essential to solve at least one 
problem in CDM galaxy formation theory), the challenge for theorists is 
to first understand the role of baryons within any viable dark matter model.

Thus, it is becoming clear that using dark matter-only simulations leads to 
biased predictions for the distribution of dark matter in galaxies. 
The recent successes in modeling the baryonic component of galaxies have
allowed theorists, {\it for the first time}, to realistically model dwarf 
galaxies \citep{Governato2010, Governato2012, Shen2013, Trujillo-Gomez2013, 
Brooks2014}.  Hence, simulators are finally in a position to be able to make 
predictions for observations that include the effect of baryons on galaxy 
evolution.

Our advances in understanding baryonic physics require that we re-evaluate 
current observational data with a new perspective.  If baryons 
alter the evolution of dark matter halos, what are the real limits of the 
currently favored models?  The goal of this review is to cast a critical 
eye on the observations in light of our favored models.  What are the 
successes and failures of the models?  What are potential paths forward to 
break degeneracies and to rule out models?

In what follows, I will assume that 100\% of the dark matter follows a 
given model.  I do not discuss the possibility of mixed 
models, (e.g., dark matter as a mix of both CDM and WDM, or WDM with 
self-interactions, or that some fraction of the dark matter is dissipative).
There are currently a few intriguing signals that may be interpreted 
as an indirect detection of dark matter.  These results will be reviewed
in Section~\ref{section2},
but much of the power of observations lies in the ability to constrain 
our favored models.  I will discuss the two popular models already 
mentioned, WDM and SIDM in Sections~\ref{section3} and \ref{section4}, 
respectively.  The current observational constraints are
already hinting that WDM cannot be warm enough to substantially 
differentiate it from CDM.  SIDM offers a more tantalizing path forward, 
and I will highlight future theoretical and observational probes to
test SIDM models.

\section{Indirect Detection}
\label{section2}

Before discussing the properties of galaxies that constrain the dark
matter model, I first discuss the evidence for a more straightforward
astrophysical signal.  Two possible paths may lead to detectable 
standard model particles indicative of the presence of dark matter.
The first path is annihilation of dark matter, and the second is decay 
of a dark matter particle. Annihilation of dark matter in the Universe 
today is an expected signal of a favored candidate for a CDM particle: 
the Weakly Interacting Massive Particle (WIMP).  In the WIMP model, dark 
matter particles are a thermal relic that ``froze out'' of equilibrium 
in the early Universe.  Freeze-out occurs when the rate of annihilation 
between dark matter particles is outpaced by the Hubble expansion 
\citep{Gondolo1991, kolbturner}.  After freeze-out, annihilation does not 
significantly decrease the WIMP number density, but will continue at low 
rates in the Universe today. Annihilation is particularly likely to happen 
in the densest regions of the Universe, i.e., in galaxies, and particularly 
in high density galaxy centers.

Despite its popularity in the theory community, it is by no means certain that
dark matter is a WIMP-like thermal relic \citep{Feng2008}. For example, if there 
is a primordial excess of WIMP particles over their anti-particle pairs (as 
there seems to have been with baryons to anti-baryons), the WIMPs and anti-WIMPs 
may continue to annihilate until nearly all the anti-particles are 
eliminated.\footnote[3]{This requires an annihilation cross-section larger than 
that of a typical CDM WIMP \citep{Buckley2011b}.}  The particle excess that is 
left over is the relic dark matter density in the Universe today. This option 
has become known as ``asymmetric dark matter''  \citep{Nussinov1985, Barr1990, 
Kaplan1992, Kribs2010, Buckley2011}.  As there is no particle to annihilate 
against today, indirect detection from annihilation is not expected in such 
models. However, a second type of indirect signal is possible if the dark matter 
particles instead decay. The lifetime of such decay must be very long 
\citep[$\gtrsim 10^{26}$ seconds][]{Ibarra2013}, but such models can be 
constructed. Decaying dark matter has even been suggested as a solution to 
CDM's small scale problems \citep[e.g.,][]{Wang2014}. 

The spectral signatures of annihilating or decaying dark matter can vary
greatly between theoretical models. Annihilation of two dark matter particles
into two photons would result in a spectral line of gamma rays with energy 
equal to the dark matter mass. Alternatively, annihilation could proceed into
Standard Model quarks, leptons, or $W/Z$ bosons, which provide a continuum
of gamma ray energies through their decays and bremsstrahlung.  Decay, on the 
other hand, is typically expected to result in a photon with an energy that 
is half of the mass of the dark matter particle, yielding an emission line at 
a specific wavelength rather than a spectrum. The morphology of any indirect 
signal can be used to distinguish the two options (annihilation or decay), as 
annihilation is proportional to the dark matter density squared, while decay 
is only proportional to the density itself.  

As of this writing, there is an exciting hint of an annihilation spectrum 
seen from our Galactic Center.  Likewise, there are also two unidentified 
lines, one at 130 GeV and the other at 3.5 keV, that are being discussed as 
possible indications of dark matter.

An excess distribution of gamma-rays from the Galactic Center has been 
seen in {\it Fermi} Gamma-Ray Space Telescope data that is not attributable 
to any known or understood source \citep{Hooper2011, Daylan2014}.  While 
the Galactic Center is a complicated place, full of baryonic physics that 
can contribute gamma-rays \citep{Abazajian2012, Boyarsky2011}, this excess 
is roughly spherical in shape and extends out to at least $\sim$10$^{\circ}$, 
and likely to several kpc \citep{Daylan2014, Hooper2013, Huang2013}.  It is 
seen after subtraction of a model for the gas disk, and of known gamma-ray 
point sources.  While originally suggested to be a population of pulsars, the 
extended distribution seems to rule out this possibility.\footnote[4]{Note  
that this excess should also not be confused with the Fermi bubbles 
\citep{Su2010}, which are likely caused by a past energetic event such as 
accretion in the Galactic Center.} The excess can be fit by a dark matter 
density distribution that follows a ``cuspy'' profile that scales as 
$\rho \propto r^{-\gamma}$, with $\gamma \sim$1.2 \citep{Daylan2014}.

Unfortunately, no other searches for excess gamma-rays due to dark matter 
annihilation have yet revealed a signal.  Despite the nearness of the 
Magellanic Clouds, they are gas-rich, making a gamma-ray signal from dark 
matter annihilation difficult to extract from the signal of cosmic rays 
interacting with the galactic interstellar medium \citep{Tasitsiomi2004, 
fermilmc}.  The most popular place to search for annihilating dark matter 
is in the dwarf spheroidal 
galaxies of the Milky Way, as they are gas-free and dark matter-dominated
\citep{Strigari2007, Strigari2008, Kuhlen2008, Kuhlen2010}.  To date, no 
significant detection has been found, and the dwarfs yield an upper limit 
that place bounds on the WIMP model \citep{Geringer2011, fermidsph, fermidsph2, 
hessdsph, hessdsph1}.  However, the
dark matter densities of the dwarf spheroidals may be too low to be 
detected with current measurements \citep{Walker2011, Cholis2012}.  A 
better hope for detection would be to find a signal from a faint, 
as yet undiscovered dwarf that happens to be relatively nearby so that the 
flux in gamma-rays is large \citep{He2013}.  Again, a search for such a signal 
in gamma-rays has not revealed any conclusive targets \citep{Buckley2010, 
Belikov2012, Hooper2012}.  The Dark Energy Survey (DES) offers the best 
hope of identifying such a dwarf in the near future, as it will be the 
first to survey the southern Galactic hemisphere for faint dwarfs 
\citep{He2013}.

{\it Fermi} data is also the source of the tentative 130 GeV line 
\citep{Weniger2012, Tempel2012}.  This line has been suggested to be an 
instrumental line \citep{Whiteson2012, Whiteson2013}, but so far no one has 
been able to conclusively demonstrate this \citep{Finkbeiner2013}.  Because 
{\it Fermi} is an all sky survey, one might expect the significance of 
this line to increase with time with more data if it is truly due to 
a dark matter source.  Instead, the significance has fluctuated 
\citep{fermi130}.  At the moment, there are no other gamma-ray telescopes 
within this energy range that can test whether the line may be instrumental. 
Given the intriguing nature of this line and the inability to rule out 
systematic effects, {\it Fermi} has recently altered its survey strategy 
to spend more time on the Galactic Center in an attempt to better understand 
whether this line is related to dark matter.  Intriguingly, $\sim$100 GeV 
has long been favored as the WIMP mass, as a WIMP model with this mass 
provides a natural fit to the relic dark matter density in the Universe 
after freeze-out \citep{Gondolo1991, kolbturner}.

As this review was being written, another possible line associated with 
decaying dark matter was observed, in the x-ray at 3.5 keV.  This line 
has been seen both in individual objects \citep[the Andromeda galaxy 
and the Perseus galaxy cluster,][]{Boyarsky2014}, and the stacked spectrum of 
73 clusters \citep{Bulbul2014}.  Most of the data comes from the {\it XMM} 
x-ray telescope, though \citet{Bulbul2014} also searched for it in 
{\it Chandra} data.  The same line was detected in the {\it Chandra} data 
for the Perseus cluster, consistent with the {\it XMM} flux, but it was not 
detected in {\it Chandra} data for the Virgo cluster.  Unlike the 130 GeV 
line, a line at 3.5 keV can be searched for with multiple current telescopes 
(both {\it Chandra} and two separate detectors on {\it XMM}, but also 
{\it Suzaku}), allowing to test if the detection is an instrumental line.  
The stacked analysis in \citet{Bulbul2014} already argues against an 
instrumental line, as the varying redshifts of the sources should wash out any 
instrumental feature.  Note that in this case, the mass of the dark matter 
particle would be 7 keV, making it a WDM particle candidate.  Sterile 
neutrinos are a popular candidate for WDM.  I will discuss in the next 
section whether a 7 keV sterile neutrino is consistent with observed 
galaxy properties.  

\section{WDM}
\label{section3}

\begin{figure*}
\center
\includegraphics[width=0.90\textwidth]{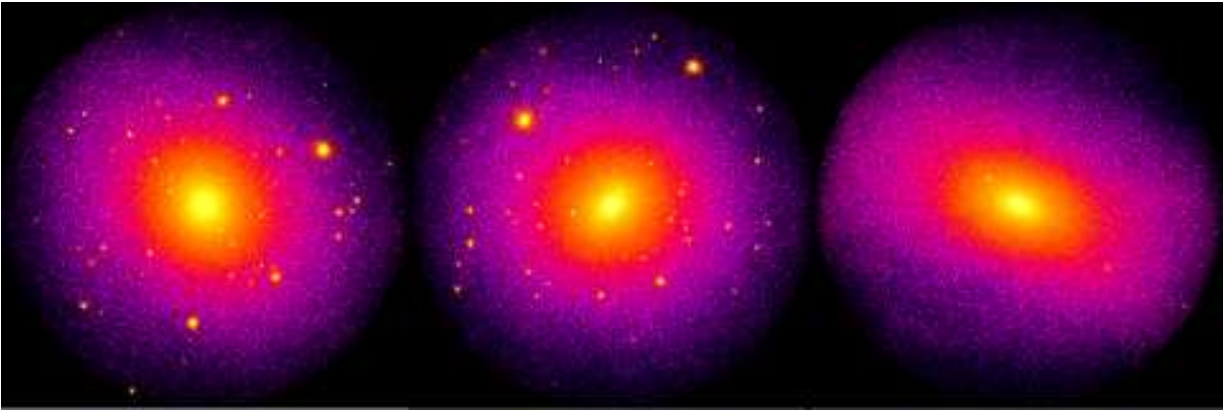}
\caption{
A 10$^{10}$ M$_{\odot}$ halo run with SIDM (left), CDM (middle), and WDM (right).
These simulations are dark matter only.  Each image is 50kpc across.  Color 
corresponds to density. The lowest densities (blue) are 100$\rho_{crit}$, and 
the highest densities (white) are 10$^6$$\rho_{crit}$.
The SIDM model (left) has been run with $\sigma = 2$ cm$^2$/g (Fry et al., in 
prep).  Note the more spherical shape of the halo compared to the CDM run, as well 
as the lower densities reached at the very center ($\sim10^5$$\rho_{crit}$).  
The WDM model (right) has been run with the power spectrum corresponding to a 
2 keV thermal relic mass.}
\label{fig1}
\end{figure*}

WDM is usually invoked to explain a lack of low mass halos that are a generic 
prediction of CDM (see Fig.~\ref{fig1}). This problem extends beyond just the 
missing satellites problem \citep{Moore1999, Klypin1999} and into the field 
\citep{Klypin2014}.  In CDM, the mass function of dark matter halos increases 
toward smaller mass halos.  Using a characteristic velocity at a given halo 
mass, the velocity function, $n(V) \propto V^{\alpha}$ rises toward small halos 
with $\alpha \sim-3$.  The HI {\sc alfalfa} survey \citep{Giovanelli2005} has allowed 
for a test of the CDM velocity function to lower masses than previous optical 
surveys.  The velocity function measured by the {\sc alfalfa} survey is much more
shallow ($\alpha \sim-0.8$) than CDM predicts \citep{Papastergis2011}.  The 
shallower slope is better described by the halo mass function predicted 
in WDM models \citep{Schneider2013}.  Likewise, a semi-analytic model of galaxy 
formation in a WDM scenario is a better fit to observed central 
and satellite luminosity functions \citep{Menci2012, Nierenberg2013}.

While CDM has power all the way down to very small scales \citep[e.g., Earth 
mass halos,][]{Anderhalden2013}, the higher streaming velocities of WDM at high 
redshift prevent it from initially collapsing into small halos with shallow 
gravitational wells 
\citep{Bode2001}.  The halo mass at which WDM can begin to gravitationally 
coalesce is set by the mass (and hence velocity) of the WDM particle.  Thus,
quantifying the amount of small scale structure in the Universe can place 
constraints on the mass of a WDM particle.  Early studies to determine 
whether dark matter was hot or cold showed that relativistic dark matter 
(e.g., neutrinos) would erase structures up to tens of Mpc, yet we see 
structure on smaller scales in the Universe \citep[and it is well described 
by the CDM power spectrum,][]{White1983}.  Given the success of CDM in 
describing the observed power on large scales \citep{Hlozek2012} while 
failing on small scales, WDM can be thought of as the Goldilocks solution.

A popular candidate for a WDM particle is the sterile neutrino, or a 
right-handed neutrino.  In the standard model, all fermions are expected 
to come in both left and right-handed varieties.  The left-handed neutrino 
participates in weak interactions, while the right-handed neutrino does 
not (hence, it is sterile), making it difficult to detect.  While the canonical
CDM candidate, the WIMP, is expected to be a thermal relic of the early 
Universe, it is very difficult to devise a scenario in which the sterile 
neutrino is a thermal relic, and an alternative scenario must be invoked  
\citep{Dodelson1994}.  However, the transfer function (the modification to 
the power spectrum) resulting from these alternative models has the same 
shape as a thermal production mechanism.  This allows the mass of the 
sterile neutrino to be directly compared to the mass of a thermal relic 
\citep{Colombi1996}.  If the source of the 3.5 keV line mentioned above  
was a thermal relic, it would have a mass in the range of 1.5 -- 3.0 keV
\citep{Abazajian2014}.  In what follows, I will quote the equivalent thermal 
relic WDM mass for comparison to CDM models.

There are multiple independent observations that can constrain the WDM 
mass, e.g., phase-space constraints \citep{Boyarsky2009, Horiuchi2014}, 
gravitational lensing \citep{Miranda2007}, satellite abundance 
\citep{Maccio2010, Polisensky2011, Anderhalden2013, Horiuchi2014}, the 
amount of small scale structure in the Lyman-$\alpha$ forest 
\citep{Viel2006, Seljak2006, Viel2008}, and the earliest epoch of star 
formation \citep{Barkana2001, Mesinger2005, deSouza2013, 
Pacucci2013}.  Some of the tightest constraints on the WDM mass come 
from observations of the Lyman-$\alpha$ forest at 2.5 $< z <$ 5.5 
interpreted using hydrodynamical simulations. For many years these 
results suggested that a WDM particle with mass $>$ 1 keV was allowed 
\citep{Viel2006, Seljak2006, Viel2008}.  However, a recent update set 
new limits on the WDM mass to be $>$ 3.3 keV at the 2$\sigma$ level 
\citep{Viel2013}.  Hence, the limits from the Lyman-$\alpha$ forest may 
be at odds with the tentative 3.5 keV x-ray line depending on the thermal 
relic equivalent mass \citep{Abazajian2014}.

\subsection{WDM as a solution to CDM's small scale problems}

The erasure of substructure in WDM has the ability to solve at least one 
of the major problems in CDM, the missing satellites problem \citep{Moore1999, 
Klypin1999}.  In recent years, a number of new satellites have been detected 
at fainter luminosities \citep{Willman2005, SatData9, SatData11, 
SatKinematics, SatData16, SatData18, SatData19}.  All of these 
``ultra-faint'' dwarfs have been detected in the {\it Sloan Digital Sky 
Survey} (SDSS).  Accounting for the footprint and magnitude limits of SDSS 
suggests that there may be hundreds of faint galaxies orbiting the Milky 
Way that remain undetected \citep{Willman2004, SatKinematics, Tollerud2008, 
Walsh2009}.
The existence of hundreds of ultra-faint galaxies requires that the mass 
of the WDM candidate would need to be greater than $\sim$2 keV 
\citep{Maccio2010, Polisensky2011, Anderhalden2013, Horiuchi2014}. 

Because the phase--space density of dark matter should never be higher 
than its initial density at decoupling, it was originally suggested that 
WDM might naturally lead to the existence of matter cores in galaxies 
\citep{tremainegunn, Dalcanton2001, Boyarsky2009}.  However, WDM cannot explain the large 
cores that we observe in galaxies (e.g., $\sim$1kpc core in a dwarf 
galaxy with stellar mass of 10$^8$ M$_{\odot}$) without violating the 
mass limits imposed by other observational constraints.  To create a 
1kpc dark matter core, the mass of the WDM particle would need to be 
$\sim$0.1 keV \citep{Maccio2012}, a low mass which is already ruled out 
by both the Lyman-$\alpha$ forest and the amount of substructure observed 
around the Milky Way.  At 2 keV, roughly the lower limit allowed by the 
abundance of substructure, the core size drops below 10pc.  Hence, if WDM is on the 
order of $\sim$2 keV, then a separate mechanism for creating dark matter 
cores in galaxies is still required.  Energetic feedback from supernovae
provides a natural mechanism for dark matter core creation within the 
allowed WDM mass range \citep{Pontzen2012,Governato2012}.

WDM has also been invoked to solve the Too Big to Fail problem found in 
the Milky Way and M31 satellites.  The central densities of halos should 
be lower in WDM models because structure formation occurs later.  In CDM, 
small structures form first, but in WDM models this smallest structure is 
wiped out, causing structure formation to be delayed compared to CDM 
\citep{Lovell2012}.  It has been established that the concentration 
of a halo is related to formation time, with earlier forming halos being 
more concentrated than later forming halos \citep{Wechsler2002, Zhao2003}.  
Hence, WDM halos are less concentrated, and there is less mass enclosed 
at a fixed radius \citep{Lovell2014}.  In the case of the dwarf spheroidals, 
the mass is measured most robustly at the half light radii, which are 
typically $\lesssim$1kpc for the luminous dwarfs \citep{McConnachie2012}. 
The mass enclosed at these small radii is sufficiently lowered to solve 
the Too Big to Fail Problem \citep{Lovell2012}, which requires masses to 
be lower by a factor of $\sim$2-4.  However, to fully solve the problem 
with no other contributing solution, the mass of the WDM particle cannot 
be larger than $\sim$2 keV \citep{Schneider2013}.  In other words, tension 
exists between the allowed mass range for WDM from the Lyman-$\alpha$ forest 
($>$ 3.3 keV) and the mass range required in order to solve the problems 
of the satellites. If the dark matter mass is indeed above $\sim$3 keV, 
then an additional process is still required to bring the masses of the 
luminous satellites in line with observations.  Fortunately, a baryonic 
solution (enhanced tidal stripping in the presence of a disk) exists 
that could solve this problem \citep{Penarrubia2010, Arraki2014, Brooks2014}. 

\subsection{Future Prospects} 

If further investigation of the 3.5 keV x-ray line proves that it is 
difficult to explain as something other than dark matter, the mass of 
this WDM particle needs to be reconciled with other observational 
constraints.  The thermal relic equivalent mass of an originating sterile 
neutrino \citep[1.5-3.0 keV,][]{Abazajian2014} is already in tension with the 
limits set by the Lyman-$\alpha$ forest, suggesting that we would need to 
re-evaluate our interpretation of the hydrodynamic simulations used to 
place the Lyman-$\alpha$ forest constraints.  Further, a 1.5 keV WDM particle 
is hard to reconcile with the number of ultra-faint halos already observed 
around the Milky Way, though $\sim$3 keV is not.  A 1.5 keV WDM particle
would suggest that we have been biased in finding faint dwarfs, so that our 
extrapolations to the full number not yet detected are overly generous.  
A more complete census of the number of ultra-faint galaxies and their 
distribution on the sky is required before this can be reconciled.  
Fortunately, there are a number of upcoming surveys ({\it Skymapper}, 
DES, and LSST) that should be able to inventory hundreds of faint satellites 
if they exist.

A liberal reading of the observational constraints suggests that a minimum 
mass of $\sim$2 keV is allowed, but a more conservative reading of the 
Lyman-$\alpha$ forest limits suggests an even heavier particle.  Even 2 keV 
is broadly consistent with the number of satellites around the Milky Way, and 
anything heavier is nearly indistinguishable from CDM in terms of the amount 
of small-scale structure formed.  In fact, assuming a WDM particle of 2 keV, 
there is very little difference in the resulting structure of any individual 
galaxy between CDM and WDM.  The concentration -- mass 
relation for WDM dark matter halos is essentially identical to CDM in 
this mass range \citep{Schneider2012}. 
When baryons are added, a slight contraction of the dark matter halo 
is seen in the CDM case compared to the WDM case \citep{Herpich2014, 
Lovell2014}.  \citet{Herpich2014} simulated Milky Way-mass and 
smaller galaxies in both CDM and WDM models that include baryons.  They 
attribute halo contraction in CDM to the existence of subhalos that 
drive disk instabilities, causing gas to flow the center of galaxies 
and leading to contraction.  By the same argument, the WDM simulations 
with baryons have less star formation at $z < 1$ due to a lack of 
subhalo induced instabilities.  
 
If the only change between the CDM+baryon and allowed WDM+baryon models 
is in slightly less concentrated galaxies and slightly lower star formation  
rates at low $z$, then it will be extremely difficult to disentangle 
WDM from baryons.   In fact, \citet{Herpich2014} and \cite{Governato2014} 
demonstrated that the resulting change in the star formation history and 
concentration of a galaxy is more sensitive to the details of star formation 
than it is to the range of allowed WDM masses. 

Given the lack of evidence for a WDM particle at low $z$, it appears 
that the best route to constrain the WDM model further is to probe the 
faintest structures at high $z$ to quantify the formation times of the 
smallest halos.  While a number of studies have already attempted to 
use high $z$ star formation to constrain the WDM particle mass 
\citep{Barkana2001, Mesinger2005, deSouza2013}, astronomers are now 
pursuing a series of observations that will allow us to probe to fainter 
structure than ever before.  These observations use lensing clusters to 
identify magnified galaxies at high $z$.  The Cluster Lensing And 
Supernovae survey with Hubble \citep[CLASH][]{Postman2012} has already 
idenfied two candidate galaxies at $z \gtrsim 10$ \citep{Zheng2012, 
Coe2013}.\footnote[5]{Note that a combinination of {\it HST} surveys 
have also identified six candidate galaxies at $z \sim10$ without 
lensing \citep{Bouwens2014}.} \citet{Pacucci2013} recently argued 
that a very high number density of halos is required to be seen in 
the small volumes that the lensing studies are probing. Two
candidate galaxies at $z \gtrsim 10$ restricts the WDM mass to be 
heavier than 1 keV.  However, the ongoing {\it Hubble Space Telescope} 
Frontier Fields observations\footnote[6]{www.stsci.edu/hst/campaigns/frontier-fields} 
will push $\sim$3 magnitudes deeper.  These lensing observations should 
put stronger bounds on the allowed WDM mass.  If additional $z \sim10$ 
galaxy candidates are identified and confirmed, WDM is likely to be ruled 
out.


\section{SIDM}
\label{section4}

SIDM is usually invoked to solve the cusp/core problem in CDM. 
Motivation behind a model for SIDM can be found in examining the 
standard model of particle physics.  Given the number of particles 
that exist, it seems natural to ask ourselves if dark matter may be 
more complicated than we tend to assume.  Is dark matter another particle with 
small interactions with the standard model (like a sterile neutrino)?  Or 
might the ``dark sector'' contain a similarly complex model that contains 
multiple particles?  In a more complex model, there may be a mediator 
particle that can be exchanged by the dark matter.  In galaxies, this 
particle exhange would occur most frequently where dark matter is most 
closely packed together, i.e., in the center of galaxies.  The exchange 
redistributes the energy of the dark matter particles.  Assuming an 
elastic scattering of particles, the overall effect is to heat the 
inner regions of galaxies so that particles move outward, transforming 
a cuspy inner density profile into a cored profile \citep{Spergel2000}.  

The redistribution of dark matter in SIDM models has several other 
observational implications in addition to core creation.  The scattering 
tends to equalize the velocity of particles, leading to a constant velocity 
dispersion profile within the scale radius of the galaxy (where 
collisions are relatively frequent), rather than an increasing profile 
as predicted by CDM \citep{Vogelsberger2012, Rocha2013}. The redistribution 
also acts to transform a triaxial dark matter halo into a more spherical 
distribution (see Fig.~\ref{fig1}).

The predicted circularization of the halo shapes led to a quick dismissal 
of the SIDM model when it was first invoked in the early 2000's. Core 
sizes observed in dwarf galaxies can be used to set constraints on the 
cross section for interactions, $\sigma$ in cm$^2$/g, in SIDM.  These 
same limits, when applied to clusters, suggested very circularized halos 
\citep{Yoshida2000}. Maps of clusters showed that they were much more 
elliptical than predicted by SIDM \citep{Miralda2002}.  Because of these 
results, SIDM was generally neglected for the following decade.  However, 
the question of halo shapes has been revisited recently.  
\citet{Peter2013} demonstrated that the $\sigma$ values required 
to match dwarf galaxies do not lead to enough change in the halo 
shapes of clusters to significantly distinguish them from CDM.

There are currently two models for SIDM being explored.  The simplest 
case posits that, no matter the relative velocities of the two dark 
matter particles, there is a constant cross-section for interaction. 
On the other hand, it is not unreasonable 
to assume that it becomes easier for dark matter particles to scatter 
as their relative velocities become smaller. Introducing a Yukawa 
potential to the model \citep{Loeb2011} leads to a velocity-dependent 
cross-section. This has the additional benefit of leading to core 
formation in dwarf galaxies, while not altering the shapes of the 
larger cluster halos where velocities are larger, and avoiding the 
earlier problems posed by a constant-velocity cross-section.

\subsection{SIDM as a solution to CDM's small scale problems}

As discussed above, SIDM is invoked to solve the cusp/core problem 
in CDM \citep{Spergel2000}.  Assuming a constant-velocity cross-section, 
a minimum $\sigma > 0.1$ cm$^2$/g is required in order to achieve the 
large cores we see in dwarf galaxies \citep{Loeb2011}.  This is bounded 
by observations on the massive end, where the shapes of cluster halos 
suggest $\sigma < 1$ cm$^2$/g \citep{Peter2013, Rocha2013, 
Vogelsberger2012}.

SIDM does not do nearly as well as WDM at solving the problems with 
the Milky Way's satellites, though.  Around the time that early studies 
of SIDM were pointing to overly spherical cluster shapes, other authors 
noted that subhalos that traveled through the dense regions of their 
parent halos should experience interactions that could lead to easier 
disruption of the satellites.  Some authors concluded that the large 
number of observed subhalos in clusters was also a strike against SIDM 
\citep{Gnedin2001}.  Again, more recent work shows that the number of 
disrupted halos is small enough that the discrepancy in subhalo numbers 
between SIDM and CDM would be hard to detect \citep{Rocha2013}. 
When comparing CDM and SIDM simulations that both neglect baryons, 
the surviving subhalo mass function for elastic scattering models in 
the range of allowed $\sigma$ values (0.1 to 1 cm$^2$/g) is identical 
\citep{Zavala2013}.  Thus, SIDM does not solve the missing satellites 
problem.

While not significantly reducing the number of satellites, it has been 
suggested that SIDM may help alleviate the Too Big to Fail problem.  
\citet{Rocha2013} suggested a constant-velocity model with 
$\sigma = 0.1$ would create large enough cores to lower the central 
densities of satellites to bring them into line with observations.  
However, this was based on an extrapolation of simulation 
results below their resolution limits.  \citet{Zavala2013}
instead showed that $\sigma = 0.1$ could not reduce the central 
masses of the satellites enough to match observations.  They 
suggested a minimum $\sigma > 0.6$ cm$^2$/g is necessary to 
alleviate the Too Big to Fail problem.  However, even this value 
would not fully explain Fornax.  As one of the brightest satellites, 
Fornax is expected to have been formed in one of the largest subhalos. 
It's observed low velocity dispersion cannot be fully fit by the 
0.6 cm$^2$/g model.  This tension suggests that even with core 
creation, some additional mechanism is still necessary to reduce the 
densities of the most luminous satellites enough to match observations. 
Again, a baryonic solution exists that could solve this problem 
\citep{Penarrubia2010, Zolotov2012, Arraki2014, Brooks2014}.

\subsection{Future Prospects} 

Initial analytic work suggests that the effect of baryons may 
substantially alter the predictions from SIDM models that neglect baryons. 
\citet{Kaplinghat2014} found that contraction of the baryons in a dark 
matter halo will shrink the size of the dark matter core formed by 
scatterings in SIDM.  For the Milky Way, the core size when neglecting 
baryons can be as large as the scale radius, $\sim$20kpc.  Contraction 
of baryons shinks the core to 0.5kpc \citep{Kaplinghat2014}. This is a 
dramatic difference that needs to be confirmed by SIDM simulations 
that include baryons.  Because supernova feedback can also lead to 
dark matter core creation \citep{Pontzen2012}, the effects of SIDM, 
contraction, and supernovae will all need to be carefully understood 
and disentangled.  

To date, \citet{Vogelsberger2014} are the only group to have 
published simulations of SIDM that include baryons. However, the 
simulations do not include baryonic feedback that leads to dark matter 
core creation, so the combined influence of core creation from both 
supernovae and SIDM scatterings has not been examined. Despite this, their 
model already demonstrates that the stellar component may be altered from 
the CDM case, suggesting that the stellar distribution of galaxies may 
allow us to probe the dark matter content.  More simulations, 
particularly with feedback that independently leads to cores, are 
necessary to explore these trends further.

It is also critical to note that all of the current bounds on $\sigma$ 
have been derived by comparing dark matter-only SIDM simulations to 
observations.  If baryons lead to a dramatic change  in the central 
regions of galaxies compared to dark matter-only SIDM models, then all 
of the current bounds will need to be re-examined.  This is particularly 
true in massive galaxies and clusters.  Clusters of galaxies with masses 
$> 10^{14}$ M$_{\odot}$ have scale radii $\sim$150 kpc.  If core size is 
comparable to the scale radius, core sizes this large are already ruled 
out \citep{Rocha2013}.  However, might baryons shrink the core size in 
clusters to an allowed size?  Recent measurements of brightest cluster  
galaxies have found evidence for cores, but on the scales of a few kpc 
to several tens of kpc \citep{Newman2013}.  It is difficult for baryonic 
physics to explain core sizes of tens of kpc.  Might these core sizes 
instead be indicative of SIDM with baryonic contraction?  Better modeling 
is required to answer this question.

Galaxies more massive than the Milky Way are dominated by baryons in their 
central regions, making it difficult to put tight constraints on the dark 
matter profile given uncertainties in removing the baryon contribution.  This 
makes low mass dwarf galaxies the more ideal place to test SIDM models, as 
they are dark matter-dominated and the complications of baryons are minimized. 
Dark matter-dominated dwarfs already outline a clear prediction to identify 
SIDM from CDM: even if baryonic physics 
can create dark matter cores in galaxies, it will do so in a distinctly 
different mass regime from SIDM. In the allowed velocity-dependent models 
of \citet{Vogelsberger2012}, or in the constant-velocity models with 
$\sigma \sim1$ cm$^2$/g, even halos as small as Draco (with stellar 
mass 3$\times$10$^5$ M$_{\odot}$) have large dark matter cores.  This 
is not true in CDM+baryon models.  The creation of a core in baryonic 
models is tied to the amount of energy that has been injected, i.e., 
to the amount of stars that have formed.  Halos the size of Draco are 
too small to have had enough star formation to create a large core 
\citep{Garrison-kimmel2013}.  The exact scaling of core size and mass 
will depend on how well stellar/supernovae feedback couples to the ISM. 
Assuming a coupling of 40\%, \citet{Penarrubia2012} showed that roughly 
10$^7$ M$_{\odot}$ in stars is necessary to create kpc-sized cores. 
Adopting this standard, it implies that kpc-sized cores cannot be 
created in halos as faint as Draco.  If such large cores were to be 
identified in these faint halos, it would be strong evidence for SIDM.  

Unfortunately, determining whether such faint halos have cores is 
a daunting observational challenge.  It has been claimed that Draco has 
both a core \citep{Wolf2012} and a cusp \citep{Jardel2013}.  The results are not 
only disparate for Draco, but even for the more massive and luminous dwarf 
Spheroidal satellites \citep{Strigari2010, Walker2011b, Hayashi2012, Jardel2012, 
Breddels2013b}. The interpretation is most sensitive to assumptions about the 
anisotropy of the stellar orbits, an unknown \citep{Evans2009, Battaglia2013, 
Richardson2013, Richardson2013b}.  

Regardless of the {\it slope} of the dark 
matter density profile in these dwarfs, there is mounting observational evidence 
that the {\it normalization} of the dark matter density is lower than predicted by 
CDM, i.e., that dwarf galaxies have lower masses than predicted within a given 
radius.  For dwarf satellites, this may be caused by tidal stripping while 
orbiting around the parent halo's disk \citep{Brooks2014}.  However, even field 
dwarf galaxies that should not have been influenced by tidal stripping seem to 
have lower masses than predicted by CDM.  Abundance matching of stellar masses 
to halo masses suggests that galaxies in the stellar mass range below 
10$^7$ M$_{\odot}$ have rotational velocities consistently lower than expected 
in CDM \citep{Ferrero2012, Garrison-Kimmel2014, Papastergis2014}.  If this trend cannot be be 
explained by baryonic physics, then again SIDM would provide a natural explanation.

\newpage
\section{Conclusions}

Baryonic phyiscs has been shown to be able to solve all of the problems 
of galaxy formation within CDM that are highlighted in this review: 
(1) the cusp/core problem \citep{Navarro1996, Read2005, deSouza2011, Governato2010, 
Pontzen2012, Teyssier2013, diCintio2014}, (2) the existence of 
bulgeless disk galaxies \citep{Governato2010, Brook2011, Teyssier2013}, 
(3) the missing satellites problem \citep{Brooks2013}, and (4) the ``Too 
Big to Fail'' problem \citep{Zolotov2012, Arraki2014, Brooks2014}.  
Only baryons have the potential to solve all of these problems together. 
Of the two alternative models to CDM that are discussed in this article, 
neither can solve these problems simultaneously.  WDM may alleviate the 
problems in the satellites, but cannot create dark matter cores.  SIDM can 
create dark matter cores, but cannot alleviate the satellite problems.  
Importantly, neither WDM nor SIDM can create bulgeless disk galaxies 
without baryonic feedback.  

Because baryons have been shown to so dramatically alter the evolution 
of the dark matter structure in the center of galaxies and in satellites, 
it is clear that dark matter-only simulations cannot be used to make accurate 
predictions on small scales.  Future preditions for galaxy formation 
{\it in any model} must consider the role of baryons.  This review 
has highlighted the future prospects of constraining two popular 
dark matter models as an alternative to CDM.

The current limits on the mass of a WDM particle are relatively heavy, 
$\gtrsim$2 keV.  At 2 keV, structure formation in WDM is nearly 
indistinguishable from CDM.  Theorists have already begun to include 
baryons in predictions for WDM, but the main difference is that less 
star formation occurs in WDM models \citep{Herpich2014, Governato2014}.  Current simulations 
are more sensitive to the star formation prescription than they are to 
the mass of the WDM particle.  Hence, identifying WDM from CDM based on 
simulation predictions requires a better understanding of star formation 
than we currently have.  Rather, the best path forward for ruling out 
or favoring WDM is through observations.  Further observations of the 
tentative 3.5 keV x-ray line, and the amount of star formation at 
$z > 10$, are currently the optimal observations to pursue.

Simulations of galaxies formed with SIDM that include baryons are needed. 
While analytic 
predictions are beginning to appear \citep{Kaplinghat2014}, their 
dramatic predictions need to be confirmed.  If baryons are as important 
as claimed in reducing SIDM core sizes, the bounds on $\sigma$ will need 
to be re-evaluated.  On the theoretical side, progress will be made 
utilizing simulations over a range of galaxy masses.  Presumably the 
scaling relations of galaxies might show systematic differences between 
SIDM and CDM, allowing the model to be constrained.  On the observational 
side, the existence of kpc-sized cores in galaxies with less than 
10$^7$ M$_{\odot}$ in stellar mass would favor SIDM models.  These 
faint galaxies already show hints of being less massive than predicted 
by CDM \citep{Ferrero2012, Papastergis2014}.  Hence, understanding the mass distributions 
in these faint field galaxies is the immediate best observational test 
of SIDM.




\begin{thebibliography}{182}
\expandafter\ifx\csname natexlab\endcsname\relax\def\natexlab#1{#1}\fi

\bibitem[{{Abazajian}(2014)}]{Abazajian2014}
{Abazajian}, K.~N. 2014, Physical Review Letters, 112, 161303

\bibitem[{{Abazajian} \& {Kaplinghat}(2012)}]{Abazajian2012}
{Abazajian}, K.~N., \& {Kaplinghat}, M. 2012, \prd, 86, 083511

\bibitem[{{Abdo} {et~al.}(2010){Abdo}, {Ackermann}, {Ajello}, {Atwood},
  {Baldini}, {Ballet}, {Barbiellini}, {Bastieri}, {Bechtol}, {Bellazzini},
  {Berenji}, {Bloom}, {Bonamente}, {Borgland}, {Bregeon}, {Brez}, {Brigida},
  {Bruel}, {Burnett}, {Buson}, {Caliandro}, {Cameron}, {Caraveo}, {Casandjian},
  {Cecchi}, {Chekhtman}, {Cheung}, {Chiang}, {Ciprini}, {Claus},
  {Cohen-Tanugi}, {Conrad}, {de Angelis}, {de Palma}, {Digel}, {Silva},
  {Drell}, {Drlica-Wagner}, {Dubois}, {Dumora}, {Farnier}, {Favuzzi}, {Fegan},
  {Focke}, {Fortin}, {Frailis}, {Fukazawa}, {Fusco}, {Gargano}, {Gehrels},
  {Germani}, {Giebels}, {Giglietto}, {Giordano}, {Glanzman}, {Godfrey},
  {Grenier}, {Grove}, {Guillemot}, {Guiriec}, {Gustafsson}, {Harding}, {Hays},
  {Horan}, {Hughes}, {Jackson}, {Jeltema}, {J{\'o}hannesson}, {Johnson},
  {Johnson}, {Johnson}, {Kamae}, {Katagiri}, {Kataoka}, {Kerr},
  {Kn{\"o}dlseder}, {Kuss}, {Lande}, {Latronico}, {Lemoine-Goumard}, {Longo},
  {Loparco}, {Lott}, {Lovellette}, {Lubrano}, {Madejski}, {Makeev},
  {Mazziotta}, {McEnery}, {Meurer}, {Michelson}, {Mitthumsiri}, {Mizuno},
  {Moiseev}, {Monte}, {Monzani}, {Moretti}, {Morselli}, {Moskalenko}, {Murgia},
  {Nolan}, {Norris}, {Nuss}, {Ohsugi}, {Omodei}, {Orlando}, {Ormes}, {Paneque},
  {Panetta}, {Parent}, {Pelassa}, {Pepe}, {Pesce-Rollins}, {Piron}, {Porter},
  {Profumo}, {Rain{\`o}}, {Rando}, {Razzano}, {Reimer}, {Reimer}, {Reposeur},
  {Ritz}, {Rodriguez}, {Roth}, {Sadrozinski}, {Sander}, {Saz Parkinson},
  {Scargle}, {Schalk}, {Sellerholm}, {Sgr{\`o}}, {Siskind}, {Smith}, {Smith},
  {Spandre}, {Spinelli}, {Strickman}, {Suson}, {Takahashi}, {Takahashi},
  {Tanaka}, {Thayer}, {Thayer}, {Thompson}, {Tibaldo}, {Torres}, {Tramacere},
  {Uchiyama}, {Usher}, {Vasileiou}, {Vilchez}, {Vitale}, {Waite}, {Wang},
  {Winer}, {Wood}, {Ylinen}, {Ziegler}, {Bullock}, {Kaplinghat}, {Martinez}, \&
  {Fermi LAT Collaboration}}]{fermidsph}
{Abdo}, A.~A. {et~al.} 2010, \apj, 712, 147

\bibitem[{{Ackermann} {et~al.}(2014){Ackermann}, {Albert}, {Anderson},
  {Baldini}, {Ballet}, {Barbiellini}, {Bastieri}, {Bechtol}, {Bellazzini},
  {Bissaldi}, {Bloom}, {Bonamente}, {Bouvier}, {Brandt}, {Bregeon}, {Brigida},
  {Bruel}, {Buehler}, {Buson}, {Caliandro}, {Cameron}, {Caragiulo}, {Caraveo},
  {Cecchi}, {Charles}, {Chekhtman}, {Chiang}, {Ciprini}, {Claus},
  {Cohen-Tanugi}, {Conrad}, {D'Ammando}, {de Angelis}, {Dermer}, {Digel}, {do
  Couto e Silva}, {Drell}, {Drlica-Wagner}, {Essig}, {Favuzzi}, {Ferrara},
  {Franckowiak}, {Fukazawa}, {Funk}, {Fusco}, {Gargano}, {Gasparrini},
  {Giglietto}, {Giroletti}, {Godfrey}, {Gomez-Vargas}, {Grenier}, {Guiriec},
  {Gustafsson}, {Hayashida}, {Hays}, {Hewitt}, {Hughes}, {Jogler}, {Kamae},
  {Kn{\"o}dlseder}, {Kocevski}, {Kuss}, {Larsson}, {Latronico}, {Llena Garde},
  {Longo}, {Loparco}, {Lovellette}, {Lubrano}, {Martinez}, {Mayer},
  {Mazziotta}, {Michelson}, {Mitthumsiri}, {Mizuno}, {Moiseev}, {Monzani},
  {Morselli}, {Moskalenko}, {Murgia}, {Nemmen}, {Nuss}, {Ohsugi}, {Orlando},
  {Ormes}, {Perkins}, {Piron}, {Pivato}, {Porter}, {Rain{\`o}}, {Rando},
  {Razzano}, {Razzaque}, {Reimer}, {Reimer}, {Ritz}, {S{\'a}nchez-Conde},
  {Sehgal}, {Sgr{\`o}}, {Siskind}, {Spinelli}, {Strigari}, {Suson}, {Tajima},
  {Takahashi}, {Thayer}, {Tibaldo}, {Tinivella}, {Torres}, {Uchiyama}, {Usher},
  {Vandenbroucke}, {Vianello}, {Vitale}, {Werner}, {Winer}, {Wood}, {Wood},
  {Zaharijas}, {Zimmer}, \& {Fermi-LAT Collaboration}}]{fermidsph2}
{Ackermann}, M. {et~al.} 2014, \prd, 89, 042001

\bibitem[{{Agertz} {et~al.}(2013){Agertz}, {Kravtsov}, {Leitner}, \&
  {Gnedin}}]{Agertz2013}
{Agertz}, O., {Kravtsov}, A.~V., {Leitner}, S.~N., \& {Gnedin}, N.~Y. 2013,
  \apj, 770, 25

\bibitem[{{Aharonian} {et~al.}(2009){Aharonian}, {Akhperjanian}, {de Almeida},
  {Bazer-Bachi}, {Behera}, {Benbow}, {Bernl{\"o}hr}, {Boisson}, {Bochow},
  {Borrel}, {Braun}, {Brion}, {Brucker}, {Brun}, {B{\"u}hler}, {Bulik},
  {B{\"u}sching}, {Boutelier}, {Carrigan}, {Chadwick}, {Charbonnier}, {Chaves},
  {Cheesebrough}, {Chounet}, {Clapson}, {Coignet}, {Costamante}, {Dalton},
  {Degrange}, {Deil}, {Dickinson}, {Djannati-Ata{\"i}}, {Domainko}, {Drury},
  {Dubois}, {Dubus}, {Dyks}, {Dyrda}, {Egberts}, {Emmanoulopoulos}, {Espigat},
  {Farnier}, {Feinstein}, {Fiasson}, {F{\"o}rster}, {Fontaine}, {F{\"u}ssling},
  {Gabici}, {Gallant}, {G{\'e}rard}, {Giebels}, {Glicenstein}, {Gl{\"u}ck},
  {Goret}, {Hadjichristidis}, {Hauser}, {Hauser}, {Heinz}, {Heinzelmann},
  {Henri}, {Hermann}, {Hinton}, {Hoffmann}, {Hofmann}, {Holleran}, {Hoppe},
  {Horns}, {Jacholkowska}, {de Jager}, {Jung}, {Katarzy{\'n}ski}, {Kaufmann},
  {Kendziorra}, {Kerschhaggl}, {Khangulyan}, {Kh{\'e}lifi}, {Keogh}, {Komin},
  {Kosack}, {Lamanna}, {Lenain}, {Lohse}, {Marandon}, {Martin},
  {Martineau-Huynh}, {Marcowith}, {Maurin}, {McComb}, {Medina}, {Moderski},
  {Moulin}, {Naumann-Godo}, {de Naurois}, {Nedbal}, {Nekrassov}, {Niemiec},
  {Nolan}, {Ohm}, {Olive}, {de O{\~n}a Wilhelmi}, {Orford}, {Osborne},
  {Ostrowski}, {Panter}, {Pedaletti}, {Pelletier}, {Petrucci}, {Pita},
  {P{\"u}hlhofer}, {Punch}, {Quirrenbach}, {Raubenheimer}, {Raue}, {Rayner},
  {Renaud}, {Rieger}, {Ripken}, {Rob}, {Rosier-Lees}, {Rowell}, {Rudak},
  {Rulten}, {Ruppel}, {Sahakian}, {Santangelo}, {Schlickeiser}, {Sch{\"o}ck},
  {Schr{\"o}der}, {Schwanke}, {Schwarzburg}, {Schwemmer}, {Shalchi}, {Skilton},
  {Sol}, {Spangler}, {Stawarz}, {Steenkamp}, {Stegmann}, {Superina}, {Tam},
  {Tavernet}, {Terrier}, {Tibolla}, {van Eldik}, {Vasileiadis}, {Venter},
  {Vialle}, {Vincent}, {Vivier}, {V{\"o}lk}, {Volpe}, {Wagner}, {Ward},
  {Zdziarski}, \& {Zech}}]{hessdsph1}
{Aharonian}, F. {et~al.} 2009, \apj, 691, 175

\bibitem[{{Anderhalden} \& {Diemand}(2013)}]{Anderhalden2013}
{Anderhalden}, D., \& {Diemand}, J. 2013, \jcap, 4, 9

\bibitem[{{Arraki} {et~al.}(2014){Arraki}, {Klypin}, {More}, \&
  {Trujillo-Gomez}}]{Arraki2014}
{Arraki}, K.~S., {Klypin}, A., {More}, S., \& {Trujillo-Gomez}, S. 2014,
  \mnras, 438, 1466

\bibitem[{{Aumer} {et~al.}(2013){Aumer}, {White}, {Naab}, \&
  {Scannapieco}}]{Aumer2013}
{Aumer}, M., {White}, S.~D.~M., {Naab}, T., \& {Scannapieco}, C. 2013, \mnras,
  434, 3142

\bibitem[{{Barkana} {et~al.}(2001){Barkana}, {Haiman}, \&
  {Ostriker}}]{Barkana2001}
{Barkana}, R., {Haiman}, Z., \& {Ostriker}, J.~P. 2001, \apj, 558, 482

\bibitem[{{Barkana} \& {Loeb}(1999)}]{Barkana1999}
{Barkana}, R., \& {Loeb}, A. 1999, \apj, 523, 54

\bibitem[{{Barnes} \& {Efstathiou}(1987)}]{Barnes1987}
{Barnes}, J., \& {Efstathiou}, G. 1987, \apj, 319, 575

\bibitem[{{Barr} {et~al.}(1990){Barr}, {Sekhar Chivukula}, \&
  {Farhi}}]{Barr1990}
{Barr}, S.~M., {Sekhar Chivukula}, R., \& {Farhi}, E. 1990, Physics Letters B,
  241, 387

\bibitem[{{Battaglia} {et~al.}(2013){Battaglia}, {Helmi}, \&
  {Breddels}}]{Battaglia2013}
{Battaglia}, G., {Helmi}, A., \& {Breddels}, M. 2013, \nar, 57, 52

\bibitem[{{Belikov} {et~al.}(2012){Belikov}, {Buckley}, \&
  {Hooper}}]{Belikov2012}
{Belikov}, A.~V., {Buckley}, M.~R., \& {Hooper}, D. 2012, \prd, 86, 043504

\bibitem[{{Belokurov} {et~al.}(2008){Belokurov}, {Walker}, {Evans}, {Faria},
  {Gilmore}, {Irwin}, {Koposov}, {Mateo}, {Olszewski}, \& {Zucker}}]{SatData16}
{Belokurov}, V. {et~al.} 2008, \apjl, 686, L83

\bibitem[{{Belokurov} {et~al.}(2010){Belokurov}, {Walker}, {Evans}, {Gilmore},
  {Irwin}, {Just}, {Koposov}, {Mateo}, {Olszewski}, {Watkins}, \&
  {Wyrzykowski}}]{SatData19}
---. 2010, \apjl, 712, L103

\bibitem[{{Bode} {et~al.}(2001){Bode}, {Ostriker}, \& {Turok}}]{Bode2001}
{Bode}, P., {Ostriker}, J.~P., \& {Turok}, N. 2001, \apj, 556, 93

\bibitem[{{Bouwens} {et~al.}(2014){Bouwens}, {Illingworth}, {Oesch}, {Trenti},
  {Labbe'}, {Bradley}, {Carollo}, {van Dokkum}, {Gonzalez}, {Holwerda},
  {Franx}, {Spitler}, {Smit}, \& {Magee}}]{Bouwens2014}
{Bouwens}, R.~J. {et~al.} 2014, ArXiv e-prints, 1403.4295

\bibitem[{{Boyarsky} {et~al.}(2011){Boyarsky}, {Malyshev}, \&
  {Ruchayskiy}}]{Boyarsky2011}
{Boyarsky}, A., {Malyshev}, D., \& {Ruchayskiy}, O. 2011, Physics Letters B,
  705, 165

\bibitem[{{Boyarsky} {et~al.}(2009){Boyarsky}, {Ruchayskiy}, \&
  {Iakubovskyi}}]{Boyarsky2009}
{Boyarsky}, A., {Ruchayskiy}, O., \& {Iakubovskyi}, D. 2009, \jcap, 3, 5

\bibitem[{{Boyarsky} {et~al.}(2014){Boyarsky}, {Ruchayskiy}, {Iakubovskyi}, \&
  {Franse}}]{Boyarsky2014}
{Boyarsky}, A., {Ruchayskiy}, O., {Iakubovskyi}, D., \& {Franse}, J. 2014,
  ArXiv e-prints, 1402.4119

\bibitem[{{Boylan-Kolchin} {et~al.}(2011){Boylan-Kolchin}, {Bullock}, \&
  {Kaplinghat}}]{Boylan-kolchin2011}
{Boylan-Kolchin}, M., {Bullock}, J.~S., \& {Kaplinghat}, M. 2011, \mnras, 415,
  L40

\bibitem[{{Boylan-Kolchin} {et~al.}(2012){Boylan-Kolchin}, {Bullock}, \&
  {Kaplinghat}}]{Boylan-kolchin2012}
---. 2012, \mnras, 2657

\bibitem[{{Breddels} \& {Helmi}(2013)}]{Breddels2013b}
{Breddels}, M.~A., \& {Helmi}, A. 2013, \aap, 558, A35

\bibitem[{{Brook} {et~al.}(2011){Brook}, {Governato}, {Ro{\v s}kar}, {Stinson},
  {Brooks}, {Wadsley}, {Quinn}, {Gibson}, {Snaith}, {Pilkington}, {House}, \&
  {Pontzen}}]{Brook2011}
{Brook}, C.~B. {et~al.} 2011, \mnras, 415, 1051

\bibitem[{{Brooks} {et~al.}(2013){Brooks}, {Kuhlen}, {Zolotov}, \&
  {Hooper}}]{Brooks2013}
{Brooks}, A.~M., {Kuhlen}, M., {Zolotov}, A., \& {Hooper}, D. 2013, \apj, 765,
  22

\bibitem[{{Brooks} \& {Zolotov}(2014)}]{Brooks2014}
{Brooks}, A.~M., \& {Zolotov}, A. 2014, \apj, 786

\bibitem[{{Buckley}(2011)}]{Buckley2011b}
{Buckley}, M.~R. 2011, \prd, 84, 043510

\bibitem[{{Buckley} \& {Hooper}(2010)}]{Buckley2010}
{Buckley}, M.~R., \& {Hooper}, D. 2010, \prd, 82, 063501

\bibitem[{{Buckley} \& {Randall}(2011)}]{Buckley2011}
{Buckley}, M.~R., \& {Randall}, L. 2011, Journal of High Energy Physics, 9, 9

\bibitem[{{Bulbul} {et~al.}(2014){Bulbul}, {Markevitch}, {Foster}, {Smith},
  {Loewenstein}, \& {Randall}}]{Bulbul2014}
{Bulbul}, E., {Markevitch}, M., {Foster}, A., {Smith}, R.~K., {Loewenstein},
  M., \& {Randall}, S.~W. 2014, \apj, 789, 13

\bibitem[{{Cholis} \& {Salucci}(2012)}]{Cholis2012}
{Cholis}, I., \& {Salucci}, P. 2012, \prd, 86, 023528

\bibitem[{{Christensen} {et~al.}(2012){Christensen}, {Quinn}, {Governato},
  {Stilp}, {Shen}, \& {Wadsley}}]{Christensen2012}
{Christensen}, C., {Quinn}, T., {Governato}, F., {Stilp}, A., {Shen}, S., \&
  {Wadsley}, J. 2012, \mnras, 425, 3058

\bibitem[{{Coe} {et~al.}(2013){Coe}, {Zitrin}, {Carrasco}, {Shu}, {Zheng},
  {Postman}, {Bradley}, {Koekemoer}, {Bouwens}, {Broadhurst}, {Monna}, {Host},
  {Moustakas}, {Ford}, {Moustakas}, {van der Wel}, {Donahue}, {Rodney},
  {Ben{\'{\i}}tez}, {Jouvel}, {Seitz}, {Kelson}, \& {Rosati}}]{Coe2013}
{Coe}, D. {et~al.} 2013, \apj, 762, 32

\bibitem[{{Collins} {et~al.}(2014){Collins}, {Chapman}, {Rich}, {Ibata},
  {Martin}, {Irwin}, {Bate}, {Lewis}, {Pe{\~n}arrubia}, {Arimoto}, {Casey},
  {Ferguson}, {Koch}, {McConnachie}, \& {Tanvir}}]{Collins2014}
{Collins}, M.~L.~M. {et~al.} 2014, \apj, 783, 7

\bibitem[{{Colombi} {et~al.}(1996){Colombi}, {Dodelson}, \&
  {Widrow}}]{Colombi1996}
{Colombi}, S., {Dodelson}, S., \& {Widrow}, L.~M. 1996, \apj, 458, 1

\bibitem[{{Dalcanton} \& {Hogan}(2001)}]{Dalcanton2001}
{Dalcanton}, J.~J., \& {Hogan}, C.~J. 2001, \apj, 561, 35

\bibitem[{{Daylan} {et~al.}(2014){Daylan}, {Finkbeiner}, {Hooper}, {Linden},
  {Portillo}, {Rodd}, \& {Slatyer}}]{Daylan2014}
{Daylan}, T., {Finkbeiner}, D.~P., {Hooper}, D., {Linden}, T., {Portillo},
  S.~K.~N., {Rodd}, N.~L., \& {Slatyer}, T.~R. 2014, ArXiv e-prints, 1402.6703

\bibitem[{{de Blok} \& {Bosma}(2002)}]{deblok2002}
{de Blok}, W.~J.~G., \& {Bosma}, A. 2002, \aap, 385, 816

\bibitem[{{de Blok} {et~al.}(2001){de Blok}, {McGaugh}, \&
  {Rubin}}]{deblok2001}
{de Blok}, W.~J.~G., {McGaugh}, S.~S., \& {Rubin}, V.~C. 2001, \aj, 122, 2396

\bibitem[{{de Blok} {et~al.}(2008){de Blok}, {Walter}, {Brinks},
  {Trachternach}, {Oh}, \& {Kennicutt}}]{deblok2008}
{de Blok}, W.~J.~G., {Walter}, F., {Brinks}, E., {Trachternach}, C., {Oh}, S.,
  \& {Kennicutt}, R.~C. 2008, \aj, 136, 2648

\bibitem[{{de Souza} {et~al.}(2013){de Souza}, {Mesinger}, {Ferrara}, {Haiman},
  {Perna}, \& {Yoshida}}]{deSouza2013}
{de Souza}, R.~S., {Mesinger}, A., {Ferrara}, A., {Haiman}, Z., {Perna}, R., \&
  {Yoshida}, N. 2013, \mnras, 432, 3218

\bibitem[{{de Souza} {et~al.}(2011){de Souza}, {Rodrigues}, {Ishida}, \&
  {Opher}}]{deSouza2011}
{de Souza}, R.~S., {Rodrigues}, L.~F.~S., {Ishida}, E.~E.~O., \& {Opher}, R.
  2011, \mnras, 415, 2969

\bibitem[{{Di Cintio} {et~al.}(2014){Di Cintio}, {Brook}, {Macci{\`o}},
  {Stinson}, {Knebe}, {Dutton}, \& {Wadsley}}]{diCintio2014}
{Di Cintio}, A., {Brook}, C.~B., {Macci{\`o}}, A.~V., {Stinson}, G.~S.,
  {Knebe}, A., {Dutton}, A.~A., \& {Wadsley}, J. 2014, \mnras, 437, 415

\bibitem[{{Di Cintio} {et~al.}(2013){Di Cintio}, {Knebe}, {Libeskind}, {Brook},
  {Yepes}, {Gottl{\"o}ber}, \& {Hoffman}}]{diCintio2013}
{Di Cintio}, A., {Knebe}, A., {Libeskind}, N.~I., {Brook}, C., {Yepes}, G.,
  {Gottl{\"o}ber}, S., \& {Hoffman}, Y. 2013, \mnras, 431, 1220

\bibitem[{{Dodelson} \& {Widrow}(1994)}]{Dodelson1994}
{Dodelson}, S., \& {Widrow}, L.~M. 1994, Physical Review Letters, 72, 17

\bibitem[{{Dutton}(2009)}]{Dutton2009b}
{Dutton}, A.~A. 2009, \mnras, 396, 121

\bibitem[{{Evans} {et~al.}(2009){Evans}, {An}, \& {Walker}}]{Evans2009}
{Evans}, N.~W., {An}, J., \& {Walker}, M.~G. 2009, \mnras, 393, L50

\bibitem[{{Feng} \& {Kumar}(2008)}]{Feng2008}
{Feng}, J.~L., \& {Kumar}, J. 2008, Physical Review Letters, 101, 231301

\bibitem[{{Fermi-LAT Collaboration}(2013)}]{fermi130}
{Fermi-LAT Collaboration}. 2013, ArXiv e-prints, 1305.5597

\bibitem[{{Ferrero} {et~al.}(2012){Ferrero}, {Abadi}, {Navarro}, {Sales}, \&
  {Gurovich}}]{Ferrero2012}
{Ferrero}, I., {Abadi}, M.~G., {Navarro}, J.~F., {Sales}, L.~V., \& {Gurovich},
  S. 2012, \mnras, 425, 2817

\bibitem[{{Finkbeiner} {et~al.}(2013){Finkbeiner}, {Su}, \&
  {Weniger}}]{Finkbeiner2013}
{Finkbeiner}, D.~P., {Su}, M., \& {Weniger}, C. 2013, \jcap, 1, 29

\bibitem[{{Garrison-Kimmel} {et~al.}(2014){Garrison-Kimmel}, {Boylan-Kolchin},
  {Bullock}, \& {Kirby}}]{Garrison-Kimmel2014}
{Garrison-Kimmel}, S., {Boylan-Kolchin}, M., {Bullock}, J.~S., \& {Kirby},
  E.~N. 2014, ArXiv e-prints, 1404.5313

\bibitem[{{Garrison-Kimmel} {et~al.}(2013){Garrison-Kimmel}, {Rocha},
  {Boylan-Kolchin}, {Bullock}, \& {Lally}}]{Garrison-kimmel2013}
{Garrison-Kimmel}, S., {Rocha}, M., {Boylan-Kolchin}, M., {Bullock}, J.~S., \&
  {Lally}, J. 2013, \mnras, 433, 3539

\bibitem[{{Gentile} {et~al.}(2007){Gentile}, {Salucci}, {Klein}, \&
  {Granato}}]{Gentile2007}
{Gentile}, G., {Salucci}, P., {Klein}, U., \& {Granato}, G.~L. 2007, \mnras,
  375, 199

\bibitem[{{Geringer-Sameth} \& {Koushiappas}(2011)}]{Geringer2011}
{Geringer-Sameth}, A., \& {Koushiappas}, S.~M. 2011, Physical Review Letters,
  107, 241303

\bibitem[{{Giovanelli} {et~al.}(2005){Giovanelli}, {Haynes}, {Kent},
  {Perillat}, {Saintonge}, {Brosch}, {Catinella}, {Hoffman}, {Stierwalt},
  {Spekkens}, {Lerner}, {Masters}, {Momjian}, {Rosenberg}, {Springob},
  {Boselli}, {Charmandaris}, {Darling}, {Davies}, {Garcia Lambas}, {Gavazzi},
  {Giovanardi}, {Hardy}, {Hunt}, {Iovino}, {Karachentsev}, {Karachentseva},
  {Koopmann}, {Marinoni}, {Minchin}, {Muller}, {Putman}, {Pantoja}, {Salzer},
  {Scodeggio}, {Skillman}, {Solanes}, {Valotto}, {van Driel}, \& {van
  Zee}}]{Giovanelli2005}
{Giovanelli}, R. {et~al.} 2005, \aj, 130, 2598

\bibitem[{{Gnedin}(2000)}]{Gnedin2000}
{Gnedin}, N.~Y. 2000, \apj, 542, 535

\bibitem[{{Gnedin} \& {Ostriker}(2001)}]{Gnedin2001}
{Gnedin}, O.~Y., \& {Ostriker}, J.~P. 2001, \apj, 561, 61

\bibitem[{{Gondolo} \& {Gelmini}(1991)}]{Gondolo1991}
{Gondolo}, P., \& {Gelmini}, G. 1991, Nuclear Physics B, 360, 145

\bibitem[{{Governato} {et~al.}(2010){Governato}, {Brook}, {Mayer}, {Brooks},
  {Rhee}, {Wadsley}, {Jonsson}, {Willman}, {Stinson}, {Quinn}, \&
  {Madau}}]{Governato2010}
{Governato}, F. {et~al.} 2010, \nat, 463, 203

\bibitem[{{Governato} {et~al.}(2014){Governato}, {Weisz}, {Pontzen}, {Loebman},
  {Reed}, {Brooks}, {Behroozi}, {Christensen}, {Madau}, {Mayer}, {Shen},
  {Walker}, {Quinn}, \& {Wadsley}}]{Governato2014}
---. 2014, ArXiv e-prints, 1407.0022

\bibitem[{{Governato} {et~al.}(2012){Governato}, {Zolotov}, {Pontzen},
  {Christensen}, {Oh}, {Brooks}, {Quinn}, {Shen}, \& {Wadsley}}]{Governato2012}
---. 2012, \mnras, 422, 1231

\bibitem[{{Guedes} {et~al.}(2011){Guedes}, {Callegari}, {Madau}, \&
  {Mayer}}]{Guedes2011}
{Guedes}, J., {Callegari}, S., {Madau}, P., \& {Mayer}, L. 2011, \apj, 742, 76

\bibitem[{{Hayashi} \& {Chiba}(2012)}]{Hayashi2012}
{Hayashi}, K., \& {Chiba}, M. 2012, \apj, 755, 145

\bibitem[{{He} {et~al.}(2013){He}, {Bechtol}, {Hearin}, \& {Hooper}}]{He2013}
{He}, C., {Bechtol}, K., {Hearin}, A.~P., \& {Hooper}, D. 2013, ArXiv e-prints,
  1309.4780

\bibitem[{{Herpich} {et~al.}(2014){Herpich}, {Stinson}, {Macci{\`o}}, {Brook},
  {Wadsley}, {Couchman}, \& {Quinn}}]{Herpich2014}
{Herpich}, J., {Stinson}, G.~S., {Macci{\`o}}, A.~V., {Brook}, C., {Wadsley},
  J., {Couchman}, H.~M.~P., \& {Quinn}, T. 2014, \mnras, 437, 293

\bibitem[{{H.E.S.S.~Collaboration} {et~al.}(2011){H.E.S.S.~Collaboration},
  {Abramowski}, {Acero}, {Aharonian}, {Akhperjanian}, {Anton}, {Barnacka},
  {Barres de Almeida}, {Bazer-Bachi}, {Becherini}, {Becker}, {Behera},
  {Bernl{\"o}hr}, {Bochow}, {Boisson}, {Bolmont}, {Bordas}, {Borrel},
  {Brucker}, {Brun}, {Brun}, {Bulik}, {B{\"u}sching}, {Carrigan}, {Casanova},
  {Cerruti}, {Chadwick}, {Charbonnier}, {Chaves}, {Cheesebrough}, {Chounet},
  {Clapson}, {Coignet}, {Conrad}, {Dalton}, {Daniel}, {Davids}, {Degrange},
  {Deil}, {Dickinson}, {Djannati-Ata{\"i}}, {Domainko}, {Drury}, {Dubois},
  {Dubus}, {Dyks}, {Dyrda}, {Egberts}, {Eger}, {Espigat}, {Fallon}, {Farnier},
  {Fegan}, {Feinstein}, {Fernandes}, {Fiasson}, {Fontaine}, {F{\"o}rster},
  {F{\"u}{\ss}ling}, {Gallant}, {Gast}, {G{\'e}rard}, {Gerbig}, {Giebels},
  {Glicenstein}, {Gl{\"u}ck}, {Goret}, {G{\"o}ring}, {Hague}, {Hampf},
  {Hauser}, {Heinz}, {Heinzelmann}, {Henri}, {Hermann}, {Hinton}, {Hoffmann},
  {Hofmann}, {Hofverberg}, {Horns}, {Jacholkowska}, {de Jager}, {Jahn},
  {Jamrozy}, {Jung}, {Kastendieck}, {Katarzy{\'n}ski}, {Katz}, {Kaufmann},
  {Keogh}, {Kerschhaggl}, {Khangulyan}, {Kh{\'e}lifi}, {Klochkov},
  {Klu{\'z}niak}, {Kneiske}, {Komin}, {Kosack}, {Kossakowski}, {Laffon},
  {Lamanna}, {Lennarz}, {Lohse}, {Lopatin}, {Lu}, {Marandon}, {Marcowith},
  {Masbou}, {Maurin}, {Maxted}, {McComb}, {Medina}, {M{\'e}hault}, {Moderski},
  {Moulin}, {Naumann}, {Naumann-Godo}, {de Naurois}, {Nedbal}, {Nekrassov},
  {Nguyen}, {Nicholas}, {Niemiec}, {Nolan}, {Ohm}, {Olive}, {de O{\~n}a
  Wilhelmi}, {Opitz}, {Ostrowski}, {Panter}, {Paz Arribas}, {Pedaletti},
  {Pelletier}, {Petrucci}, {Pita}, {P{\"u}hlhofer}, {Punch}, {Quirrenbach},
  {Raue}, {Rayner}, {Reimer}, {Reimer}, {Renaud}, {de Los Reyes}, {Rieger},
  {Ripken}, {Rob}, {Rosier-Lees}, {Rowell}, {Rudak}, {Rulten}, {Ruppel},
  {Ryde}, {Sahakian}, {Santangelo}, {Schlickeiser}, {Sch{\"o}ck},
  {Sch{\"o}nwald}, {Schwanke}, {Schwarzburg}, {Schwemmer}, {Shalchi}, {Sikora},
  {Skilton}, {Sol}, {Spengler}, {Stawarz}, {Steenkamp}, {Stegmann}, {Stinzing},
  {Sushch}, {Szostek}, {Tavernet}, {Terrier}, {Tibolla}, {Tluczykont},
  {Valerius}, {van Eldik}, {Vasileiadis}, {Venter}, {Vialle}, {Viana},
  {Vincent}, {Vivier}, {V{\"o}lk}, {Volpe}, {Vorobiov}, {Vorster}, {Wagner},
  {Ward}, {Wierzcholska}, {Zajczyk}, {Zdziarski}, {Zech}, {Zechlin}, \&
  {H.E.S.S.~Collaboration}}]{hessdsph}
{H.E.S.S.~Collaboration} {et~al.} 2011, Astroparticle Physics, 34, 608

\bibitem[{{Hinshaw} {et~al.}(2013){Hinshaw}, {Larson}, {Komatsu}, {Spergel},
  {Bennett}, {Dunkley}, {Nolta}, {Halpern}, {Hill}, {Odegard}, {Page}, {Smith},
  {Weiland}, {Gold}, {Jarosik}, {Kogut}, {Limon}, {Meyer}, {Tucker}, {Wollack},
  \& {Wright}}]{WMAP9}
{Hinshaw}, G. {et~al.} 2013, \apjs, 208, 19

\bibitem[{{Hlozek} {et~al.}(2012){Hlozek}, {Dunkley}, {Addison}, {Appel},
  {Bond}, {Sofia Carvalho}, {Das}, {Devlin}, {D{\"u}nner}, {Essinger-Hileman},
  {Fowler}, {Gallardo}, {Hajian}, {Halpern}, {Hasselfield}, {Hilton}, {Hincks},
  {Hughes}, {Irwin}, {Klein}, {Kosowsky}, {Marriage}, {Marsden}, {Menanteau},
  {Moodley}, {Niemack}, {Nolta}, {Page}, {Parker}, {Partridge}, {Rojas},
  {Sehgal}, {Sherwin}, {Sievers}, {Spergel}, {Staggs}, {Swetz}, {Switzer},
  {Thornton}, \& {Wollack}}]{Hlozek2012}
{Hlozek}, R. {et~al.} 2012, \apj, 749, 90

\bibitem[{{Hooper} \& {Linden}(2011)}]{Hooper2011}
{Hooper}, D., \& {Linden}, T. 2011, \prd, 84, 123005

\bibitem[{{Hooper} \& {Linden}(2012)}]{Hooper2012}
---. 2012, \prd, 86, 083532

\bibitem[{{Hooper} \& {Slatyer}(2013)}]{Hooper2013}
{Hooper}, D., \& {Slatyer}, T.~R. 2013, Physics of the Dark Universe, 2, 118

\bibitem[{{Hopkins} {et~al.}(2013){Hopkins}, {Keres}, {Onorbe},
  {Faucher-Giguere}, {Quataert}, {Murray}, \& {Bullock}}]{Hopkins2013}
{Hopkins}, P.~F., {Keres}, D., {Onorbe}, J., {Faucher-Giguere}, C.-A.,
  {Quataert}, E., {Murray}, N., \& {Bullock}, J.~S. 2013, ArXiv e-prints,
  1311.2073

\bibitem[{{Horiuchi} {et~al.}(2014){Horiuchi}, {Humphrey}, {O{\~n}orbe},
  {Abazajian}, {Kaplinghat}, \& {Garrison-Kimmel}}]{Horiuchi2014}
{Horiuchi}, S., {Humphrey}, P.~J., {O{\~n}orbe}, J., {Abazajian}, K.~N.,
  {Kaplinghat}, M., \& {Garrison-Kimmel}, S. 2014, \prd, 89, 025017

\bibitem[{{Huang} {et~al.}(2013){Huang}, {Urbano}, \& {Xue}}]{Huang2013}
{Huang}, W.-C., {Urbano}, A., \& {Xue}, W. 2013, ArXiv e-prints, 1307.6862

\bibitem[{{Ibarra} {et~al.}(2013){Ibarra}, {Tran}, \& {Weniger}}]{Ibarra2013}
{Ibarra}, A., {Tran}, D., \& {Weniger}, C. 2013, International Journal of
  Modern Physics A, 28, 30040

\bibitem[{{Irwin} {et~al.}(2007){Irwin}, {Belokurov}, {Evans}, {Ryan-Weber},
  {de Jong}, {Koposov}, {Zucker}, {Hodgkin}, {Gilmore}, {Prema}, {Hebb},
  {Begum}, {Fellhauer}, {Hewett}, {Kennicutt}, {Wilkinson}, {Bramich},
  {Vidrih}, {Rix}, {Beers}, {Barentine}, {Brewington}, {Harvanek},
  {Krzesinski}, {Long}, {Nitta}, \& {Snedden}}]{SatData9}
{Irwin}, M.~J. {et~al.} 2007, \apjl, 656, L13

\bibitem[{{Jardel} \& {Gebhardt}(2012)}]{Jardel2012}
{Jardel}, J.~R., \& {Gebhardt}, K. 2012, \apj, 746, 89

\bibitem[{{Jardel} {et~al.}(2013){Jardel}, {Gebhardt}, {Fabricius}, {Drory}, \&
  {Williams}}]{Jardel2013}
{Jardel}, J.~R., {Gebhardt}, K., {Fabricius}, M.~H., {Drory}, N., \&
  {Williams}, M.~J. 2013, \apj, 763, 91

\bibitem[{{Kaplan}(1992)}]{Kaplan1992}
{Kaplan}, D.~B. 1992, Physical Review Letters, 68, 741

\bibitem[{{Kaplinghat} {et~al.}(2013){Kaplinghat}, {Keeley}, {Linden}, \&
  {Yu}}]{Kaplinghat2014}
{Kaplinghat}, M., {Keeley}, R.~E., {Linden}, T., \& {Yu}, H.-B. 2013, ArXiv
  e-prints, 1311.6524

\bibitem[{{Klypin} {et~al.}(2014){Klypin}, {Karachentsev}, {Makarov}, \&
  {Nasonova}}]{Klypin2014}
{Klypin}, A., {Karachentsev}, I., {Makarov}, D., \& {Nasonova}, O. 2014, ArXiv
  e-prints, 1405.4523

\bibitem[{{Klypin} {et~al.}(1999){Klypin}, {Kravtsov}, {Valenzuela}, \&
  {Prada}}]{Klypin1999}
{Klypin}, A., {Kravtsov}, A.~V., {Valenzuela}, O., \& {Prada}, F. 1999, \apj,
  522, 82

\bibitem[{{Kolb} \& {Turner}(1994)}]{kolbturner}
{Kolb}, E.~W., \& {Turner}, M.~S. 1994, {The early universe.} (Westview Press)

\bibitem[{{Kribs} {et~al.}(2010){Kribs}, {Roy}, {Terning}, \&
  {Zurek}}]{Kribs2010}
{Kribs}, G.~D., {Roy}, T.~S., {Terning}, J., \& {Zurek}, K.~M. 2010, \prd, 81,
  095001

\bibitem[{{Kuhlen}(2010)}]{Kuhlen2010}
{Kuhlen}, M. 2010, Advances in Astronomy, 2010

\bibitem[{{Kuhlen} {et~al.}(2008){Kuhlen}, {Diemand}, \& {Madau}}]{Kuhlen2008}
{Kuhlen}, M., {Diemand}, J., \& {Madau}, P. 2008, \apj, 686, 262

\bibitem[{{Kuzio de Naray} {et~al.}(2006){Kuzio de Naray}, {McGaugh}, {de
  Blok}, \& {Bosma}}]{Kuzio2006}
{Kuzio de Naray}, R., {McGaugh}, S.~S., {de Blok}, W.~J.~G., \& {Bosma}, A.
  2006, \apjs, 165, 461

\bibitem[{{Liu} {et~al.}(2008){Liu}, {Hu}, {Newberg}, \& {Zhao}}]{SatData11}
{Liu}, C., {Hu}, J., {Newberg}, H., \& {Zhao}, Y. 2008, \aap, 477, 139

\bibitem[{{Loeb} \& {Weiner}(2011)}]{Loeb2011}
{Loeb}, A., \& {Weiner}, N. 2011, Physical Review Letters, 106, 171302

\bibitem[{{Lovell} {et~al.}(2012){Lovell}, {Eke}, {Frenk}, {Gao}, {Jenkins},
  {Theuns}, {Wang}, {White}, {Boyarsky}, \& {Ruchayskiy}}]{Lovell2012}
{Lovell}, M.~R. {et~al.} 2012, \mnras, 420, 2318

\bibitem[{{Lovell} {et~al.}(2014){Lovell}, {Frenk}, {Eke}, {Jenkins}, {Gao}, \&
  {Theuns}}]{Lovell2014}
{Lovell}, M.~R., {Frenk}, C.~S., {Eke}, V.~R., {Jenkins}, A., {Gao}, L., \&
  {Theuns}, T. 2014, \mnras, 439, 300

\bibitem[{{Macci{\`o}} \& {Fontanot}(2010)}]{Maccio2010}
{Macci{\`o}}, A.~V., \& {Fontanot}, F. 2010, \mnras, 404, L16

\bibitem[{{Macci{\`o}} {et~al.}(2012){Macci{\`o}}, {Paduroiu}, {Anderhalden},
  {Schneider}, \& {Moore}}]{Maccio2012}
{Macci{\`o}}, A.~V., {Paduroiu}, S., {Anderhalden}, D., {Schneider}, A., \&
  {Moore}, B. 2012, \mnras, 424, 1105

\bibitem[{{McConnachie}(2012)}]{McConnachie2012}
{McConnachie}, A.~W. 2012, \aj, 144, 4

\bibitem[{{Menci} {et~al.}(2012){Menci}, {Fiore}, \& {Lamastra}}]{Menci2012}
{Menci}, N., {Fiore}, F., \& {Lamastra}, A. 2012, \mnras, 421, 2384

\bibitem[{{Mesinger} {et~al.}(2005){Mesinger}, {Perna}, \&
  {Haiman}}]{Mesinger2005}
{Mesinger}, A., {Perna}, R., \& {Haiman}, Z. 2005, \apj, 623, 1

\bibitem[{{Miralda-Escud{\'e}}(2002)}]{Miralda2002}
{Miralda-Escud{\'e}}, J. 2002, \apj, 564, 60

\bibitem[{{Miranda} \& {Macci{\`o}}(2007)}]{Miranda2007}
{Miranda}, M., \& {Macci{\`o}}, A.~V. 2007, \mnras, 382, 1225

\bibitem[{{Moore} {et~al.}(1999){Moore}, {Ghigna}, {Governato}, {Lake},
  {Quinn}, {Stadel}, \& {Tozzi}}]{Moore1999}
{Moore}, B., {Ghigna}, S., {Governato}, F., {Lake}, G., {Quinn}, T., {Stadel},
  J., \& {Tozzi}, P. 1999, \apjl, 524, L19

\bibitem[{{Navarro} {et~al.}(1996){Navarro}, {Eke}, \& {Frenk}}]{Navarro1996}
{Navarro}, J.~F., {Eke}, V.~R., \& {Frenk}, C.~S. 1996, \mnras, 283, L72

\bibitem[{{Navarro} {et~al.}(1997){Navarro}, {Frenk}, \&
  {White}}]{Navarro1997a}
{Navarro}, J.~F., {Frenk}, C.~S., \& {White}, S.~D.~M. 1997, \apj, 490, 493

\bibitem[{{Navarro} {et~al.}(2010){Navarro}, {Ludlow}, {Springel}, {Wang},
  {Vogelsberger}, {White}, {Jenkins}, {Frenk}, \& {Helmi}}]{Navarro2010}
{Navarro}, J.~F. {et~al.} 2010, \mnras, 402, 21

\bibitem[{{Newman} {et~al.}(2013){Newman}, {Treu}, {Ellis}, \&
  {Sand}}]{Newman2013}
{Newman}, A.~B., {Treu}, T., {Ellis}, R.~S., \& {Sand}, D.~J. 2013, \apj, 765,
  25

\bibitem[{{Nierenberg} {et~al.}(2013){Nierenberg}, {Treu}, {Menci}, {Lu}, \&
  {Wang}}]{Nierenberg2013}
{Nierenberg}, A.~M., {Treu}, T., {Menci}, N., {Lu}, Y., \& {Wang}, W. 2013,
  \apj, 772, 146

\bibitem[{{Nussinov}(1985)}]{Nussinov1985}
{Nussinov}, S. 1985, Physics Letters B, 165, 55

\bibitem[{{Oh} {et~al.}(2011){Oh}, {Brook}, {Governato}, {Brinks}, {Mayer}, {de
  Blok}, {Brooks}, \& {Walter}}]{Oh2011}
{Oh}, S.-H., {Brook}, C., {Governato}, F., {Brinks}, E., {Mayer}, L., {de
  Blok}, W.~J.~G., {Brooks}, A., \& {Walter}, F. 2011, \aj, 142, 24

\bibitem[{{Okamoto} {et~al.}(2008){Okamoto}, {Gao}, \& {Theuns}}]{Okamoto2008}
{Okamoto}, T., {Gao}, L., \& {Theuns}, T. 2008, \mnras, 390, 920

\bibitem[{{Pacucci} {et~al.}(2013){Pacucci}, {Mesinger}, \&
  {Haiman}}]{Pacucci2013}
{Pacucci}, F., {Mesinger}, A., \& {Haiman}, Z. 2013, \mnras, 435, L53

\bibitem[{{Papastergis} {et~al.}(2014){Papastergis}, {Giovanelli}, {Haynes}, \&
  {Shankar}}]{Papastergis2014}
{Papastergis}, E., {Giovanelli}, R., {Haynes}, M.~P., \& {Shankar}, F. 2014,
  ArXiv e-prints, 1407.4665

\bibitem[{{Papastergis} {et~al.}(2011){Papastergis}, {Martin}, {Giovanelli}, \&
  {Haynes}}]{Papastergis2011}
{Papastergis}, E., {Martin}, A.~M., {Giovanelli}, R., \& {Haynes}, M.~P. 2011,
  \apj, 739, 38

\bibitem[{{Pe{\~n}arrubia} {et~al.}(2010){Pe{\~n}arrubia}, {Benson}, {Walker},
  {Gilmore}, {McConnachie}, \& {Mayer}}]{Penarrubia2010}
{Pe{\~n}arrubia}, J., {Benson}, A.~J., {Walker}, M.~G., {Gilmore}, G.,
  {McConnachie}, A.~W., \& {Mayer}, L. 2010, \mnras, 406, 1290

\bibitem[{{Pe{\~n}arrubia} {et~al.}(2012){Pe{\~n}arrubia}, {Pontzen}, {Walker},
  \& {Koposov}}]{Penarrubia2012}
{Pe{\~n}arrubia}, J., {Pontzen}, A., {Walker}, M.~G., \& {Koposov}, S.~E. 2012,
  \apjl, 759, L42

\bibitem[{{Peebles}(1969)}]{Peebles1969}
{Peebles}, P.~J.~E. 1969, \apj, 155, 393

\bibitem[{{Peter} {et~al.}(2013){Peter}, {Rocha}, {Bullock}, \&
  {Kaplinghat}}]{Peter2013}
{Peter}, A.~H.~G., {Rocha}, M., {Bullock}, J.~S., \& {Kaplinghat}, M. 2013,
  \mnras, 430, 105

\bibitem[{{Planck Collaboration} {et~al.}(2013){Planck Collaboration}, {Ade},
  {Aghanim}, {Armitage-Caplan}, {Arnaud}, {Ashdown}, {Atrio-Barandela},
  {Aumont}, {Baccigalupi}, {Banday}, \& et~al.}]{Planck2013}
{Planck Collaboration} {et~al.} 2013, ArXiv e-prints, 1303.5076

\bibitem[{{Polisensky} \& {Ricotti}(2011)}]{Polisensky2011}
{Polisensky}, E., \& {Ricotti}, M. 2011, \prd, 83, 043506

\bibitem[{{Pontzen} \& {Governato}(2012)}]{Pontzen2012}
{Pontzen}, A., \& {Governato}, F. 2012, \mnras, 421, 3464

\bibitem[{{Pontzen} \& {Governato}(2014)}]{Pontzen2014}
---. 2014, \nat, 506, 171

\bibitem[{{Postman} {et~al.}(2012){Postman}, {Coe}, {Ben{\'{\i}}tez},
  {Bradley}, {Broadhurst}, {Donahue}, {Ford}, {Graur}, {Graves}, {Jouvel},
  {Koekemoer}, {Lemze}, {Medezinski}, {Molino}, {Moustakas}, {Ogaz}, {Riess},
  {Rodney}, {Rosati}, {Umetsu}, {Zheng}, {Zitrin}, {Bartelmann}, {Bouwens},
  {Czakon}, {Golwala}, {Host}, {Infante}, {Jha}, {Jimenez-Teja}, {Kelson},
  {Lahav}, {Lazkoz}, {Maoz}, {McCully}, {Melchior}, {Meneghetti}, {Merten},
  {Moustakas}, {Nonino}, {Patel}, {Reg{\"o}s}, {Sayers}, {Seitz}, \& {Van der
  Wel}}]{Postman2012}
{Postman}, M. {et~al.} 2012, \apjs, 199, 25

\bibitem[{{Primack}(2012)}]{PrimackAnnalen}
{Primack}, J.~R. 2012, Annalen der Physik, 524, 535

\bibitem[{{Quinn} \& {Binney}(1992)}]{Quinn1992}
{Quinn}, T., \& {Binney}, J. 1992, \mnras, 255, 729

\bibitem[{{Quinn} {et~al.}(1996){Quinn}, {Katz}, \& {Efstathiou}}]{Quinn1996}
{Quinn}, T., {Katz}, N., \& {Efstathiou}, G. 1996, \mnras, 278, L49

\bibitem[{{Read} \& {Gilmore}(2005)}]{Read2005}
{Read}, J.~I., \& {Gilmore}, G. 2005, \mnras, 356, 107

\bibitem[{{Richardson} \& {Fairbairn}(2013{\natexlab{a}})}]{Richardson2013b}
{Richardson}, T., \& {Fairbairn}, M. 2013{\natexlab{a}}, \mnras, 432, 3361

\bibitem[{{Richardson} \& {Fairbairn}(2013{\natexlab{b}})}]{Richardson2013}
---. 2013{\natexlab{b}}, ArXiv e-prints, 1305.0670

\bibitem[{{Rocha} {et~al.}(2013){Rocha}, {Peter}, {Bullock}, {Kaplinghat},
  {Garrison-Kimmel}, {O{\~n}orbe}, \& {Moustakas}}]{Rocha2013}
{Rocha}, M., {Peter}, A.~H.~G., {Bullock}, J.~S., {Kaplinghat}, M.,
  {Garrison-Kimmel}, S., {O{\~n}orbe}, J., \& {Moustakas}, L.~A. 2013, \mnras,
  430, 81

\bibitem[{{Schneider} {et~al.}(2014){Schneider}, {Anderhalden}, {Macci{\`o}},
  \& {Diemand}}]{Schneider2013}
{Schneider}, A., {Anderhalden}, D., {Macci{\`o}}, A.~V., \& {Diemand}, J. 2014,
  \mnras, 441, L6

\bibitem[{{Schneider} {et~al.}(2012){Schneider}, {Smith}, {Macci{\`o}}, \&
  {Moore}}]{Schneider2012}
{Schneider}, A., {Smith}, R.~E., {Macci{\`o}}, A.~V., \& {Moore}, B. 2012,
  \mnras, 424, 684

\bibitem[{{Seljak} {et~al.}(2006){Seljak}, {Makarov}, {McDonald}, \&
  {Trac}}]{Seljak2006}
{Seljak}, U., {Makarov}, A., {McDonald}, P., \& {Trac}, H. 2006, Physical
  Review Letters, 97, 191303

\bibitem[{{Shen} {et~al.}(2013){Shen}, {Madau}, {Conroy}, {Governato}, \&
  {Mayer}}]{Shen2013}
{Shen}, S., {Madau}, P., {Conroy}, C., {Governato}, F., \& {Mayer}, L. 2013,
  ArXiv e-prints, 1308.4131

\bibitem[{{Simon} {et~al.}(2003){Simon}, {Bolatto}, {Leroy}, \&
  {Blitz}}]{Simon2003}
{Simon}, J.~D., {Bolatto}, A.~D., {Leroy}, A., \& {Blitz}, L. 2003, \apj, 596,
  957

\bibitem[{{Simon} \& {Geha}(2007)}]{SatKinematics}
{Simon}, J.~D., \& {Geha}, M. 2007, \apj, 670, 313

\bibitem[{{Spano} {et~al.}(2008){Spano}, {Marcelin}, {Amram}, {Carignan},
  {Epinat}, \& {Hernandez}}]{Spano2008}
{Spano}, M., {Marcelin}, M., {Amram}, P., {Carignan}, C., {Epinat}, B., \&
  {Hernandez}, O. 2008, \mnras, 383, 297

\bibitem[{{Spergel} \& {Steinhardt}(2000)}]{Spergel2000}
{Spergel}, D.~N., \& {Steinhardt}, P.~J. 2000, Physical Review Letters, 84,
  3760

\bibitem[{{Springel} {et~al.}(2008){Springel}, {Wang}, {Vogelsberger},
  {Ludlow}, {Jenkins}, {Helmi}, {Navarro}, {Frenk}, \& {White}}]{Springel2008}
{Springel}, V. {et~al.} 2008, \mnras, 391, 1685

\bibitem[{{Strigari} {et~al.}(2010){Strigari}, {Frenk}, \&
  {White}}]{Strigari2010}
{Strigari}, L.~E., {Frenk}, C.~S., \& {White}, S.~D.~M. 2010, \mnras, 408, 2364

\bibitem[{{Strigari} {et~al.}(2007){Strigari}, {Koushiappas}, {Bullock}, \&
  {Kaplinghat}}]{Strigari2007}
{Strigari}, L.~E., {Koushiappas}, S.~M., {Bullock}, J.~S., \& {Kaplinghat}, M.
  2007, \prd, 75, 083526

\bibitem[{{Strigari} {et~al.}(2008){Strigari}, {Koushiappas}, {Bullock},
  {Kaplinghat}, {Simon}, {Geha}, \& {Willman}}]{Strigari2008}
{Strigari}, L.~E., {Koushiappas}, S.~M., {Bullock}, J.~S., {Kaplinghat}, M.,
  {Simon}, J.~D., {Geha}, M., \& {Willman}, B. 2008, \apj, 678, 614

\bibitem[{{Su} {et~al.}(2010){Su}, {Slatyer}, \& {Finkbeiner}}]{Su2010}
{Su}, M., {Slatyer}, T.~R., \& {Finkbeiner}, D.~P. 2010, \apj, 724, 1044

\bibitem[{{Swaters} {et~al.}(2003){Swaters}, {Madore}, {van den Bosch}, \&
  {Balcells}}]{Swaters2003}
{Swaters}, R.~A., {Madore}, B.~F., {van den Bosch}, F.~C., \& {Balcells}, M.
  2003, \apj, 583, 732

\bibitem[{{Tasitsiomi} {et~al.}(2004){Tasitsiomi}, {Gaskins}, \&
  {Olinto}}]{Tasitsiomi2004}
{Tasitsiomi}, A., {Gaskins}, J., \& {Olinto}, A.~V. 2004, Astroparticle
  Physics, 21, 637

\bibitem[{{Tempel} {et~al.}(2012){Tempel}, {Hektor}, \& {Raidal}}]{Tempel2012}
{Tempel}, E., {Hektor}, A., \& {Raidal}, M. 2012, \jcap, 9, 32

\bibitem[{{Teyssier} {et~al.}(2013){Teyssier}, {Pontzen}, {Dubois}, \&
  {Read}}]{Teyssier2013}
{Teyssier}, R., {Pontzen}, A., {Dubois}, Y., \& {Read}, J.~I. 2013, \mnras,
  429, 3068

\bibitem[{{The Fermi/LAT collaboration} \& {Abdo}(2010)}]{fermilmc}
{The Fermi/LAT collaboration}, \& {Abdo}, A.~A. 2010, ArXiv e-prints, 1001.3298

\bibitem[{{Thoul} \& {Weinberg}(1996)}]{Thoul1996}
{Thoul}, A.~A., \& {Weinberg}, D.~H. 1996, \apj, 465, 608

\bibitem[{{Tollerud} {et~al.}(2012){Tollerud}, {Beaton}, {Geha}, {Bullock},
  {Guhathakurta}, {Kalirai}, {Majewski}, {Kirby}, {Gilbert}, {Yniguez},
  {Patterson}, {Ostheimer}, {Cooke}, {Dorman}, {Choudhury}, \&
  {Cooper}}]{Tollerud2012}
{Tollerud}, E.~J. {et~al.} 2012, \apj, 752, 45

\bibitem[{{Tollerud} {et~al.}(2008){Tollerud}, {Bullock}, {Strigari}, \&
  {Willman}}]{Tollerud2008}
{Tollerud}, E.~J., {Bullock}, J.~S., {Strigari}, L.~E., \& {Willman}, B. 2008,
  \apj, 688, 277

\bibitem[{{Trachternach} {et~al.}(2008){Trachternach}, {de Blok}, {Walter},
  {Brinks}, \& {Kennicutt}}]{Trachternach2008}
{Trachternach}, C., {de Blok}, W.~J.~G., {Walter}, F., {Brinks}, E., \&
  {Kennicutt}, Jr., R.~C. 2008, \aj, 136, 2720

\bibitem[{{Tremaine} \& {Gunn}(1979)}]{tremainegunn}
{Tremaine}, S., \& {Gunn}, J.~E. 1979, Physical Review Letters, 42, 407

\bibitem[{{Trujillo-Gomez} {et~al.}(2013){Trujillo-Gomez}, {Klypin}, {Colin},
  {Ceverino}, {Arraki}, \& {Primack}}]{Trujillo-Gomez2013}
{Trujillo-Gomez}, S., {Klypin}, A., {Colin}, P., {Ceverino}, D., {Arraki}, K.,
  \& {Primack}, J. 2013, ArXiv e-prints, 1311.2910

\bibitem[{{van den Bosch} {et~al.}(2002){van den Bosch}, {Abel}, {Croft},
  {Hernquist}, \& {White}}]{vdBosch2002}
{van den Bosch}, F.~C., {Abel}, T., {Croft}, R.~A.~C., {Hernquist}, L., \&
  {White}, S.~D.~M. 2002, \apj, 576, 21

\bibitem[{{van den Bosch} {et~al.}(2001){van den Bosch}, {Burkert}, \&
  {Swaters}}]{vdBosch2001}
{van den Bosch}, F.~C., {Burkert}, A., \& {Swaters}, R.~A. 2001, \mnras, 326,
  1205

\bibitem[{{van den Bosch} {et~al.}(2000){van den Bosch}, {Robertson},
  {Dalcanton}, \& {de Blok}}]{vandenbosch2000}
{van den Bosch}, F.~C., {Robertson}, B.~E., {Dalcanton}, J.~J., \& {de Blok},
  W.~J.~G. 2000, \aj, 119, 1579

\bibitem[{{Veilleux} {et~al.}(2005){Veilleux}, {Cecil}, \&
  {Bland-Hawthorn}}]{Veilleux2005}
{Veilleux}, S., {Cecil}, G., \& {Bland-Hawthorn}, J. 2005, \araa, 43, 769

\bibitem[{{Viel} {et~al.}(2013){Viel}, {Becker}, {Bolton}, \&
  {Haehnelt}}]{Viel2013}
{Viel}, M., {Becker}, G.~D., {Bolton}, J.~S., \& {Haehnelt}, M.~G. 2013, \prd,
  88, 043502

\bibitem[{{Viel} {et~al.}(2008){Viel}, {Becker}, {Bolton}, {Haehnelt}, {Rauch},
  \& {Sargent}}]{Viel2008}
{Viel}, M., {Becker}, G.~D., {Bolton}, J.~S., {Haehnelt}, M.~G., {Rauch}, M.,
  \& {Sargent}, W.~L.~W. 2008, Physical Review Letters, 100, 041304

\bibitem[{{Viel} {et~al.}(2006){Viel}, {Lesgourgues}, {Haehnelt}, {Matarrese},
  \& {Riotto}}]{Viel2006}
{Viel}, M., {Lesgourgues}, J., {Haehnelt}, M.~G., {Matarrese}, S., \& {Riotto},
  A. 2006, Physical Review Letters, 97, 071301

\bibitem[{{Vogelsberger} {et~al.}(2012){Vogelsberger}, {Zavala}, \&
  {Loeb}}]{Vogelsberger2012}
{Vogelsberger}, M., {Zavala}, J., \& {Loeb}, A. 2012, \mnras, 423, 3740

\bibitem[{{Vogelsberger} {et~al.}(2014){Vogelsberger}, {Zavala}, {Simpson}, \&
  {Jenkins}}]{Vogelsberger2014}
{Vogelsberger}, M., {Zavala}, J., {Simpson}, C., \& {Jenkins}, A. 2014, ArXiv
  e-prints, 1405.5216

\bibitem[{{Walker} {et~al.}(2011){Walker}, {Combet}, {Hinton}, {Maurin}, \&
  {Wilkinson}}]{Walker2011}
{Walker}, M.~G., {Combet}, C., {Hinton}, J.~A., {Maurin}, D., \& {Wilkinson},
  M.~I. 2011, \apjl, 733, L46

\bibitem[{{Walker} \& {Pe{\~n}arrubia}(2011)}]{Walker2011b}
{Walker}, M.~G., \& {Pe{\~n}arrubia}, J. 2011, \apj, 742, 20

\bibitem[{{Walsh} {et~al.}(2009){Walsh}, {Willman}, \& {Jerjen}}]{Walsh2009}
{Walsh}, S.~M., {Willman}, B., \& {Jerjen}, H. 2009, \aj, 137, 450

\bibitem[{{Wang} {et~al.}(2014){Wang}, {Peter}, {Strigari}, {Zentner}, {Arant},
  {Garrison-Kimmel}, \& {Rocha}}]{Wang2014}
{Wang}, M.-Y., {Peter}, A.~H.~G., {Strigari}, L.~E., {Zentner}, A.~R., {Arant},
  B., {Garrison-Kimmel}, S., \& {Rocha}, M. 2014, ArXiv e-prints, 1406.0527

\bibitem[{{Watkins} {et~al.}(2009){Watkins}, {Evans}, {Belokurov}, {Smith},
  {Hewett}, {Bramich}, {Gilmore}, {Irwin}, {Vidrih}, {Wyrzykowski}, \&
  {Zucker}}]{SatData18}
{Watkins}, L.~L. {et~al.} 2009, \mnras, 398, 1757

\bibitem[{{Wechsler} {et~al.}(2002){Wechsler}, {Bullock}, {Primack},
  {Kravtsov}, \& {Dekel}}]{Wechsler2002}
{Wechsler}, R.~H., {Bullock}, J.~S., {Primack}, J.~R., {Kravtsov}, A.~V., \&
  {Dekel}, A. 2002, \apj, 568, 52

\bibitem[{{Weldrake} {et~al.}(2003){Weldrake}, {de Blok}, \&
  {Walter}}]{Weldrake2003}
{Weldrake}, D.~T.~F., {de Blok}, W.~J.~G., \& {Walter}, F. 2003, \mnras, 340,
  12

\bibitem[{{Weniger}(2012)}]{Weniger2012}
{Weniger}, C. 2012, \jcap, 8, 7

\bibitem[{{White}(1984)}]{White1984}
{White}, S.~D.~M. 1984, \apj, 286, 38

\bibitem[{{White} {et~al.}(1983){White}, {Frenk}, \& {Davis}}]{White1983}
{White}, S.~D.~M., {Frenk}, C.~S., \& {Davis}, M. 1983, \apjl, 274, L1

\bibitem[{{Whiteson}(2012)}]{Whiteson2012}
{Whiteson}, D. 2012, \jcap, 11, 8

\bibitem[{{Whiteson}(2013)}]{Whiteson2013}
---. 2013, \prd, 88, 023530

\bibitem[{{Willman} {et~al.}(2005){Willman}, {Dalcanton}, {Martinez-Delgado},
  {West}, {Blanton}, {Hogg}, {Barentine}, {Brewington}, {Harvanek}, {Kleinman},
  {Krzesinski}, {Long}, {Neilsen}, {Nitta}, \& {Snedden}}]{Willman2005}
{Willman}, B. {et~al.} 2005, \apjl, 626, L85

\bibitem[{{Willman} {et~al.}(2004){Willman}, {Governato}, {Dalcanton}, {Reed},
  \& {Quinn}}]{Willman2004}
{Willman}, B., {Governato}, F., {Dalcanton}, J.~J., {Reed}, D., \& {Quinn}, T.
  2004, \mnras, 353, 639

\bibitem[{{Wolf} \& {Bullock}(2012)}]{Wolf2012}
{Wolf}, J., \& {Bullock}, J.~S. 2012, ArXiv e-prints, 1203.4240

\bibitem[{{Yoshida} {et~al.}(2000){Yoshida}, {Springel}, {White}, \&
  {Tormen}}]{Yoshida2000}
{Yoshida}, N., {Springel}, V., {White}, S.~D.~M., \& {Tormen}, G. 2000, \apjl,
  544, L87

\bibitem[{{Zavala} {et~al.}(2013){Zavala}, {Vogelsberger}, \&
  {Walker}}]{Zavala2013}
{Zavala}, J., {Vogelsberger}, M., \& {Walker}, M.~G. 2013, \mnras, 431, L20

\bibitem[{{Zhao} {et~al.}(2003){Zhao}, {Jing}, {Mo}, \&
  {B{\"o}rner}}]{Zhao2003}
{Zhao}, D.~H., {Jing}, Y.~P., {Mo}, H.~J., \& {B{\"o}rner}, G. 2003, \apjl,
  597, L9

\bibitem[{{Zheng} {et~al.}(2012){Zheng}, {Postman}, {Zitrin}, {Moustakas},
  {Shu}, {Jouvel}, {H{\o}st}, {Molino}, {Bradley}, {Coe}, {Moustakas},
  {Carrasco}, {Ford}, {Ben{\'{\i}}tez}, {Lauer}, {Seitz}, {Bouwens},
  {Koekemoer}, {Medezinski}, {Bartelmann}, {Broadhurst}, {Donahue}, {Grillo},
  {Infante}, {Jha}, {Kelson}, {Lahav}, {Lemze}, {Melchior}, {Meneghetti},
  {Merten}, {Nonino}, {Ogaz}, {Rosati}, {Umetsu}, \& {van der Wel}}]{Zheng2012}
{Zheng}, W. {et~al.} 2012, \nat, 489, 406

\bibitem[{{Zolotov} {et~al.}(2012){Zolotov}, {Brooks}, {Willman}, {Governato},
  {Pontzen}, {Christensen}, {Dekel}, {Quinn}, {Shen}, \&
  {Wadsley}}]{Zolotov2012}
{Zolotov}, A. {et~al.} 2012, \apj, 761, 71

\end{thebibliography}

\end{document}